\documentclass[10pt,aps,prl,twocolumn,superscriptaddress,amsmath,amssymb,footinbib]{revtex4-2}

\usepackage{graphicx}
\usepackage{dcolumn}
\usepackage{bm}
\usepackage[T1]{fontenc}
\usepackage{mathtools}
\usepackage[dvipsnames]{xcolor}
\usepackage{hyperref}
\hypersetup{
    colorlinks,
    linkcolor={violet},
    citecolor={ForestGreen},
    urlcolor={teal}
}
\usepackage{orcidlink}

\usepackage{dsfont}
\usepackage{tikz}
\DeclareSymbolFont{bbold}{U}{bbold}{m}{n}
\DeclareSymbolFontAlphabet{\mathbbold}{bbold}
\newcommand{\bbPhi}{\mathbbold{\Phi}}

\newcommand{\F}{\mathbb F}
\newcommand{\Sk}{\mathbb S}
\newcommand{\RF}{\mathfrak R\mathbb F}
\newcommand{\RS}{\mathfrak R\mathbb S}
\newcommand{\hb}{\hat b}
\newcommand{\hP}{\hat P}
\newcommand{\hh}{\hat h}

\newcommand{\sm}[4]{\left[\begin{smallmatrix}#1&#2\\#3&#4\end{smallmatrix}\right]}

\newcommand{\SMs}{\hyperlink{sup_mat}{Supplemental Material}}

\begin{document}

\title{Boundary $c\leq1$ Liouville CFT from the Bootstrap}

\author{Panos Betzios\orcidlink{0000-0002-5350-9404}}
\email{panos.betzios@ugent.be}
\affiliation{\href{https://www.ugent.be/we/physics-astronomy/en}{Department of Physics and Astronomy},
Ghent University, \\ Krijgslaan 281-S9, 9000 Gent, Belgium}

\author{Paul Ghiringhelli\orcidlink{0009-0003-3991-3149}}
\email{paul.ghiringhelli@polytechnique.edu}
\affiliation{\href{https://www.cpht.polytechnique.fr/?q=en}{CPHT, CNRS, \'Ecole polytechnique, Institut Polytechnique de Paris}, \\ 91120 Palaiseau, France}

\author{Olga Papadoulaki\orcidlink{0000-0001-5302-2930}}
\email{olga.papadoulaki@polytechnique.edu}
\affiliation{\href{https://www.cpht.polytechnique.fr/?q=en}{CPHT, CNRS, \'Ecole polytechnique, Institut Polytechnique de Paris}, \\ 91120 Palaiseau, France}

\author{Ioannis Tsiares\orcidlink{0000-0001-9474-4665}}
\email{ioannis.tsiares@phys.ens.fr}
\affiliation{\href{https://www.lpens.ens.psl.eu/}{Laboratoire de Physique de l'\'Ecole Normale Sup\'erieure}, ENS, Universit\'e PSL, CNRS, \\ Sorbonne Universit\'e, Universit\'e Paris Cit\'e, 24 rue Lhomond, 75005 Paris, France}

\date{\today}

\begin{abstract}\noindent
We construct boundary Liouville conformal field theory at central charge $c\leq1$ with FZZT-type boundary conditions and show that it satisfies the Cardy--Lewellen sewing constraints on the disk and the annulus. Using the analytic bootstrap we derive the so far missing boundary three-point and bulk-to-boundary structure constants and a consistent disk one-point function as explicit meromorphic functions of the Liouville momenta. In sharp contrast to $c\geq25$, the open-string spectrum between two ``timelike'' FZZT boundaries is \emph{discrete}, yet non-trivially consistent through open-closed duality with the \emph{continuous} closed-string spectrum of Ribault and Santachiara.
\end{abstract}

\maketitle

\section{Introduction}

Liouville theory stands out among two-dimensional quantum field theories as an interacting, non-compact, yet exactly solvable conformal field theory (CFT). Introduced by Polyakov as the quantum theory of the conformal factor in non-critical string theory~\cite{Polyakov:1981rd}, it governs two-dimensional quantum gravity~\cite{Knizhnik:1988ak, David:1988hj,Distler:1988jt,Ginsparg:1993is}, is non-trivially related to the quantization of Teichm\"uller space~\cite{Verlinde:1989ua,Teschner:2003at} and to non-compact quantum groups~\cite{Ponsot:1999uf,Ponsot:2000mt}, is dual to four-dimensional $\mathcal N=2$ gauge theories~\cite{Alday:2009aq}, captures universal sectors of irrational CFTs~\cite{Collier:2018exn,Collier:2019weq}, and admits a rigorous probabilistic construction~\cite{David:2014aha,Kupiainen:2017eaw,Guillarmou:2020wbo}.

What makes Liouville theory unique among 2d CFTs is that all of its data are determined, in essence, by Virasoro representation theory alone: the celebrated DOZZ formula~\cite{Dorn:1994xn,Zamolodchikov:1995aa} for its structure constants is fixed by analyticity in the Liouville momenta together with the constraints from degenerate representations~\cite{Teschner:1995yf,Teschner:2001rv}. Parametrizing the central charge as $c=13+6(b^2+b^{-2})$, one distinguishes three domains: (i) $c\geq25$, with $b$ real; (ii) $c\leq1$, with $b=i\hat b$ purely imaginary; and (iii) $c\in\mathbb C\setminus\big((-\infty,1]\cup[25,\infty)\big)$, with $b$ complex. In (i) and (ii) $b^2\in\mathbb R$, so the shifts by $b$ and $b^{-1}$ in the degenerate-field equations lie on one line and, for irrational $b^2$, generate a dense subgroup of it, which makes the solutions unique~\cite{Teschner:1995yf}. However, the solutions in these two domains are \emph{not} analytic continuations of one another~\cite{Zamolodchikov:2005fy}.

The ``spacelike'' theory with $c\geq25$ is by now thoroughly understood. In the bulk, the DOZZ structure constant together with the continuous spectrum $h\geq\tfrac{c-1}{24}$ solves crossing symmetry on the sphere and modular covariance on the torus~\cite{Ponsot:1999uf,Teschner:2001rv,Hadasz:2009sw}. On surfaces with boundaries, the one-parameter FZZT family~\cite{Fateev:2000ik,Teschner:2000md} and the discrete ZZ family~\cite{Zamolodchikov:2001ah}, the bulk--boundary~\cite{Hosomichi:2001xc, Ponsot:2003ss,Numasawa:2022cni} and boundary three-point~\cite{Ponsot:2001ng} structure constants, and their mutual consistency through the Cardy--Lewellen sewing constraints~\cite{Cardy:1991tv,Lewellen:1991tb} are on solid footing; see~\cite{Nakayama:2004vk,Ribault:2014hia} for reviews, and~\cite{Remy:2020suk,Ang:2021ird,Wu:2022zbc, Guillarmou:2024mwf,Guillarmou:2026pqt} for the probabilistic counterpart. Notably, for suitable boundary parameters the open-string spectrum between two FZZT boundaries coincides with the bulk \emph{continuum} $h\geq\frac{c-1}{24}$~\cite{Fateev:2000ik,Teschner:2000md}, whereas ZZ boundaries support only finitely many degenerate open-string states~\cite{Zamolodchikov:2001ah}. This will be our main point of contrast with the $c\leq1$ theory constructed below.

The ``timelike'' theory with $c\leq1$ is a genuinely different CFT. It first appeared in the description of rolling tachyons and S-branes~\cite{Strominger:2003fn,Schomerus:2003vv,Gutperle:2003xf}, and its bulk three-point function~\cite{Zamolodchikov:2005fy,Kostov:2005kk,Schomerus:2003vv,McElgin:2007ak,Giribet:2011zx,Harlow:2011ny} is \emph{not} the analytic continuation of DOZZ. As Ribault and Santachiara showed~\cite{Ribault:2015sxa}, a consistent bulk theory requires essentially the same continuous spectrum $h\geq\frac{c-1}{24}$ as in the spacelike case; crossing symmetry then holds numerically~\cite{Ribault:2015sxa} and, in some cases, analytically~\cite{Iorgov:2013uoa, Ribault:2023vqs,Roussillon:2025tmv}, even though the theory is non-unitary and the operator of conformal weight zero does not correspond to the identity operator with the usual fusion rule~\cite{Zamolodchikov:2005fy}. 
The theory has found applications in 2d percolation~\cite{Delfino:2010xm}, critical loop models~\cite{Picco:2016ilr,Ikhlef:2015eua} and two-dimensional de Sitter quantum gravity~\cite{Bautista:2019jau,Anninos:2021ene,Anninos:2024iwf}, and progress has been made toward its probabilistic construction~\cite{Guillarmou:2023exh,Usciati:2025cdn}.
  
On surfaces with boundaries, the situation at $c\leq1$ is very different: more than two decades after the ZZ and FZZT solutions at $c\geq25$, existing results provide only partial boundary data. Early studies considered boundary states in the rolling-tachyon and $c=1$ settings~\cite{Gutperle:2003xf,Fredenhagen:2003ut,Fredenhagen:2004cj}. For FZZT-like boundary conditions, bulk one-point and boundary two-point functions were proposed in~\cite{Bautista:2021ogd}, and an alternative one-point function was subsequently proposed using the Coulomb-gas formalism in~\cite{Giribet:2026gao}; see also~\cite{Anninos:2025fer} for a path-integral perspective.
In the closely related critical loop models, Ref.~\cite{Downing:2025wer} identified a one-parameter family of ``reflecting'' boundaries analogous to FZZT branes and determined their disk one- and two-point functions and annulus spectra. ZZ-like boundary conditions were also proposed in the Virasoro Minimal String on the basis of the cylinder amplitude~\cite{Collier:2023cyw}. In all these constructions, however, the bulk--boundary and boundary three-point structure constants remain undetermined, so none provides a complete set of boundary data whose consistency has been tested against the Cardy--Lewellen sewing constraints~\cite{Cardy:1991tv,Lewellen:1991tb}. The open-string spectra of putative ``timelike'' FZZT and ZZ boundaries likewise remain unclear; indeed, given that the bulk spectrum is a continuum, it is not even clear what to expect for the boundary channel.

We answer these questions in this Letter. Using the bootstrap (and therefore making no reference to a particular action of the theory) we construct a one-parameter family of boundary conditions which we identify as the $c\leq1$ analogue of the FZZT branes, and we show that the resulting data satisfy the Cardy--Lewellen constraints on the disk and the annulus. In particular, we derive explicit expressions for the bulk--boundary and boundary three-point structure constants, and give a concrete proposal for the boundary spectrum.

Our main results for the structure constants, in a particular normalization which we refer to as ``bootstrap normalization'', are summarized in Table~\ref{tab:data}. In the process of understanding the $c\leq1$ case, we found it useful to first construct an analogous normalization for the $c\geq25$ bulk and boundary data with usual FZZT boundaries (middle column of Table~\ref{tab:data}). It makes manifest the fact that every structure constant in $c\geq25$ Liouville CFT is a Virasoro crossing kernel---a fusion kernel $\F$ of sphere four-point blocks or a modular kernel $\Sk$ of torus one-point blocks---and every boundary sewing constraint is a Moore--Seiberg (polynomial) relation obeyed by the kernels~\footnote{Similar relations hold between the boundary structure constants and the fusion/modular \textit{matrices} in the A-series Virasoro Minimal Models~\cite{Runkel:1998he}. It is in this sense that Liouville theory is sometimes viewed as the non-compact analogue of these minimal models.}. For $c\leq1$ something similar is true with some caveats; the novel results are presented in the rightmost column of the table. We next explain the various boundary quantities and their implications, before illustrating our main method for deriving the results.

\section{The \texorpdfstring{$c\leq1$}{c<=1} boundary data and spectrum}

\paragraph*{Notations \& Definitions.} For $c\geq25$ \emph{bulk} Liouville CFT we write $c=1+6Q^2$, $Q=b+b^{-1}$, $h=\bar{h}=\frac{Q^2}4-P^2$, with $b\in(0,1]$, and the physical spectrum is at $P\in i\mathbb R_{+}$.  For $c\le1$ \emph{bulk} Liouville CFT, $b=i\hb$ with $\hb\in(0,1]$, $c=1-6(\hb-\hb^{-1})^2$, $\hh=\bar{\hh}=\frac{c-1}{24}-\hP^2$, and the spectrum is at $\hP\in\epsilon+i\mathbb R$, $\epsilon\neq0$~\cite{Ribault:2015sxa}.  Degenerate momenta at $c\geq25$ and $c\leq1$ are accordingly at $P_{(r,s)}=\frac12(rb+sb^{-1})$ and $\hP_{(r,s)}=\frac i2(r\hb-s\hb^{-1})$, $r,s\in\mathbb{Z}_{\geq1}$; we use the same formulas for negative indices, e.g.\ $\hP_{(r,-s)}=\frac i2(r\hb+s\hb^{-1})$. For $c\geq25$ sphere four-point blocks are denoted as $\mathcal V_z(P_s;P_{1234})=z^{h_s-h_1-h_2}\left(1+ O(z)\right)$ (operators at $0,z,1,\infty$) and torus one-point blocks as $\mathcal F_\tau(P;P_0)=q^{h-\frac{c}{24}}\left(1+O(q)\right)$, $q\equiv e^{2\pi i \tau}$. Virasoro characters are $\chi_P(\tau)=q^{-P^2}/\eta(\tau)$. The corresponding blocks and characters at $c\leq1$ are denoted by a hat in the obvious places. From that, we define the Liouville fusion $\F$ and modular $\Sk$ kernels for $b\in(0,1]$ via
\begin{align}
 &\mathcal V_z(P_s;P_{1234})=\int_{i\mathbb{R}}\tfrac{dP_t}{i}\F^{(b)}_{P_s,P_t}\!\sm{P_2}{P_3}{P_1}{P_4}\!\mathcal V_{1-z}(P_t;P_{3214}),\nonumber\\
 &(-i\tau)^{h_0}\mathcal F_\tau(P_s;P_0)=\int_{i\mathbb{R}}\tfrac{dP_t}{i}\Sk^{(b)}_{P_s,P_t}[P_0]\mathcal F_{-1/\tau}(P_t;P_0).
 \label{eq:kernels}
\end{align}\noindent
The explicit expressions for these kernels were first derived in~\cite{Ponsot:1999uf,Ponsot:2000mt,Teschner:2003at,Teschner:2012em,Teschner:2013tqy} (see \SMs); they are characterized as solutions of the appropriate Moore--Seiberg difference equations. They involve even, meromorphic functions of the Liouville momenta $P_i$, are invariant under $b\rightarrow b^{-1}$, and valid in general in the extended domain $b^2\in\mathbb{C}\backslash(-\infty,0]$. Their vacuum rows and columns ($P_{\mathds{1}}=P_{(1,1)}$, corresponding to $h=0$) are especially simple and read for example $\Sk^{(b)}_{P_s,P_t}[\mathds 1]=\sqrt2\cos(4\pi P_sP_t)$, $\Sk^{(b)}_{\mathds{1},P}[\mathds{1}]=\rho^{(b)}_0(P)/2$, with $\rho^{(b)}_0(P)\equiv-4\sqrt2\sin(2\pi bP)\sin(2\pi P/b)$. For the fusion kernel, $\F^{(b)}_{\mathds{1},P_1}\sm{P_2}{P_3}{P_2}{P_3}=\frac12\rho^{(b)}_0(P_1)C^{(b)}_0(P_1,P_2,P_3)$, with $C_0$ the DOZZ three-point function normalized as in~\cite{Collier:2019weq,Collier:2023cyw} (see \SMs).  For $c\leq1$ ($\hb\in(0,1]$), Ribault and one of the authors~\cite{Ribault:2023vqs} demonstrated that the same Moore--Seiberg difference equations that govern the crossing kernels yield a set of \emph{unphysical}, albeit \emph{unique} meromorphic, solutions for the modular and fusion kernels which are given by
\begin{align}
 \RF^{(\hb)}_{\hP_s,\hP_t}\!\!\sm{\hP_2}{\hP_3}{\hP_1}{\hP_4}&:=\tfrac{\hP_t}{\hP_s}\F^{(\hb)}_{i\hP_t,i\hP_s}\!\!\sm{i\hP_2}{i\hP_1}{i\hP_3}{i\hP_4},\nonumber\\
 \RS^{(\hb)}_{\hP_s,\hP_t}[\hP_0]&:=\tfrac{\hP_t}{\hP_s}\Sk^{(\hb)}_{i\hP_t,i\hP_s}[i\hP_0].
 \label{eq:VWR}
\end{align}
Here $\F$ and $\Sk$ on the RHS are the $c\ge25$ kernels evaluated at the real parameter $\hb$. $\RF$ and $\RS$ inherit all the analytic properties from $\F,\Sk$ and hence, importantly, they are \emph{odd} functions of the internal momenta $\hat{P}_s,\hat{P}_t$. The authors of \cite{Ribault:2023vqs} called the maps $\F\to\RF$, $\Sk\to\RS$ the ``Virasoro--Wick rotation'' (VWR). For later use, we also define $\hat \rho_0(P)\equiv P^2/(2\rho^{(\hat{b})}_0(iP))$ and $\hat{C}_0(P_1,P_2,P_3)\equiv1/C^{(\hb)}_0(iP_1,iP_2,iP_3)$. One of the main points of our work is that, even though the kernels (\ref{eq:VWR}) are unphysical and do \emph{not} implement crossing of the corresponding $c\leq1$ blocks, they serve as structure constants of boundary Liouville theory at $c\leq1$.

\paragraph*{The meromorphic $c\leq1$ structure constants.}With these definitions in place and the structure constants as in Table~\ref{tab:data}, we are able to verify (both analytically and numerically) the following Cardy--Lewellen sewing constraints at $c\leq1$.
\par\noindent\textit{i) Disk (4-bdy) four-point crossing symmetry} (boundary fields $\Psi_j$ at $0,\eta,1,\infty$, with boundary condition $\sigma_{j+1}$ between $\Psi_j$ and $\Psi_{j+1}$):
\begin{equation}\label{eq:CLdisk4pt}
\begin{aligned}
&
\int_{i\mathbb R+\epsilon}\!\frac{d\hat P_s}{i}
\frac{(\hat C_0)^{\sigma_4\sigma_3\sigma_1}_{\hat P_4\hat P_3\hat P_s}\,(\hat C_0)^{\sigma_3\sigma_2\sigma_1}_{\hat P_s\hat P_2\hat P_1}}{\hat S_0(\hat P_s;\hat P_{\sigma_1},\hat P_{\sigma_3})}
\hat{\mathcal V}_{\eta}(\hat P_s;\hat P_{1234})\\
&=
\int_{i\mathbb R+\epsilon}\!\frac{d\hat P_t}{i}
\frac{(\hat C_0)^{\sigma_4\sigma_2\sigma_1}_{\hat P_4\hat P_t\hat P_1}\,(\hat C_0)^{\sigma_4\sigma_3\sigma_2}_{\hat P_t\hat P_3\hat P_2}}{\hat S_0(\hat P_t;\hat P_{\sigma_2},\hat P_{\sigma_4})}
\hat{\mathcal V}_{1-\eta}(\hat P_t;\hat P_{3214}).
\end{aligned}
\end{equation}
\par\noindent\textit{ii) Disk (1-bulk, 2-bdy) three-point crossing symmetry}:
\begin{equation}\label{eq:CLdisk3pt}
\begin{aligned}
&\phi_{\eta}\!\!\int_{i\mathbb R+\epsilon}\!\frac{d\hat P_s}{i}
\frac{e^{-\frac{i\pi}{2}\hat h_s}\hat R_0(\hat P_\alpha,\hat P_s|\hat P_{\sigma_2})\,(\hat C_0)^{\sigma_2\sigma_2\sigma_1}_{\hat P_2\hat P_s\hat P_1}}{\hat S_0(\hat P_s;\hat P_{\sigma_2},\hat P_{\sigma_2})}\\
&\qquad\times\hat{\mathcal V}_{\eta}(\hat P_s;\hat P_{\alpha\alpha12})
\\
&=
\phi'_{\eta'}\!\!\int_{i\mathbb R+\epsilon}\!\frac{d\hat P'_s}{i}
\frac{e^{\frac{i\pi}{2}\hat h_{s'}}\hat R_0(\hat P_\alpha,\hat P'_s|\hat P_{\sigma_1})\,(\hat C_0)^{\sigma_2\sigma_1\sigma_1}_{\hat P_2\hat P_1\hat P'_s}}{\hat S_0(\hat P'_s;\hat P_{\sigma_1},\hat P_{\sigma_1})}\\
&\qquad\times\hat{\mathcal V}_{\eta'}(\hat P'_s;\hat P_{\alpha\alpha12}).
\end{aligned}
\end{equation}
Here $V_{\hat P_\alpha}$ is inserted at $e^{i\varphi}$ in the upper half-plane and $\Psi_{1,2}$ at $0,\infty$, with $\sigma_2$ on the positive and $\sigma_1$ on the negative real axis; $\eta=\eta'=1-e^{-2i\varphi}$, where the unprimed (primed) side is continued from $\varphi=0$ ($\varphi=\pi$), i.e.\ from opposite sides of the cut of the blocks at $\eta=1$,
$\phi_{\eta}\equiv(1-\eta)^{\hat p}e^{i\pi \hat h_\alpha}$, $\phi'_{\eta'}\equiv(1-\eta')^{\hat p}e^{-i\pi \hat h_\alpha}$, and $\hat{p}\equiv \hat h_\alpha +\frac{\hat h_1-\hat h_2}{2}$.
\par\noindent\textit{iii) Disk (2-bulk) two-point crossing symmetry}:
\begin{equation}\label{eq:CLdisk2pt}
\begin{aligned}
&\int_{i\mathbb R+\epsilon}\!\frac{d\hat P_{\rm o}}{i}
\frac{\hat R_0(\hat P_{\alpha_1},\hat P_{\rm o}|\hat P_\sigma)\,\hat R_0(\hat P_{\alpha_2},\hat P_{\rm o}|\hat P_\sigma)}{\hat S_0(\hat P_{\rm o};\hat P_\sigma,\hat P_\sigma)}\\
&\qquad\times\hat{\mathcal V}_{x}(\hat P_{\rm o};\hat P_{\alpha_1\alpha_1\alpha_2\alpha_2})\\
&=
\int_{i\mathbb R+\epsilon}\!\frac{d\hat P_{\rm c}}{i}
\hat\rho_0(\hat P_{\rm c})\hat C_0(\hat P_{\alpha_1},\hat P_{\alpha_2},\hat P_{\rm c})\hat R_0^{(\text{1-pt})}(\hat P_{\rm c}|\hat P_\sigma)\\
&\qquad\times\hat{\mathcal V}_{1-x}(\hat P_{\rm c};\hat P_{\alpha_2\alpha_1\alpha_1\alpha_2}),
\end{aligned}
\end{equation}
where $x=4\,\mathrm{Im}\,z_1\,\mathrm{Im}\,z_2/|z_1-\bar z_2|^2$ is the cross-ratio of the bulk insertions $z_{1,2}$ and their mirror images; the subscripts ${\rm o}$ and ${\rm c}$ label open- and closed-channel momenta.
\par\noindent\textit{iv) Annulus (1-bdy) one-point modular covariance}:
\begin{equation}\label{eq:CLann1pt}
\begin{aligned}
&(-i\tau)^{\hat h_\beta}\int_{i\mathbb R+\epsilon}\!\frac{d\hat P_{\rm o}}{i}\;
\frac{(\hat C_0)^{\sigma_1\sigma_1\sigma_2}_{\hat P_{\rm o}\hat P_\beta\hat P_{\rm o}}}{\hat S_0(\hat P_{\rm o};\hat P_{\sigma_1},\hat P_{\sigma_2})}\;
\hat{\mathcal F}_\tau(\hat P_{\rm o};\hat P_\beta)\\
&\;=\;
\!\int_{i\mathbb R+\epsilon}\!\frac{d\hat P_{\rm c}}{i}\,
\hat\rho_0(\hat P_{\rm c})\hat R_0(\hat P_{\rm c},\hat P_\beta|\hat P_{\sigma_1})\,\hat R_0^{(\text{1-pt})}(\hat P_{\rm c}|\hat P_{\sigma_2})\\
&\qquad\times
\hat{\mathcal F}_{-1/\tau}(\hat P_{\rm c};\hat P_\beta).
\end{aligned}
\end{equation}
\par\noindent\textit{v) Annulus partition function open-closed duality}:
\begin{equation}\label{eq:CLann0pt}
\begin{aligned}
&\sum_{e_1,e_2=\pm}\ \sum_{k,l\ge0}\hat\chi_{\mathsf P^{e_1e_2}_{(k,l)}}(\tau)=
\int_{i\mathbb R+\epsilon}\!\frac{d\hat P_{\rm c}}{i}\;
\hat\rho_0(\hat P_{\rm c})\\
&\qquad\times\hat R_0^{(\text{1-pt})}(\hat P_{\rm c}|\hat P_{\sigma_1})\,\hat R_0^{(\text{1-pt})}(\hat P_{\rm c}|\hat P_{\sigma_2})\,
\hat\chi_{\hat P_{\rm c}}(-1/\tau).
\end{aligned}
\end{equation}
Two points about \eqref{eq:CLann0pt} merit emphasis. First, the equality between its discrete sum and continuous integral follows analytically (\SMs), identifying the support $\mathsf P^{e_1e_2}_{(k,l)}$ in~\eqref{eq:Hhatopen} that will recur in all open-channel expansions. Second, it is \emph{not} the limit of (\ref{eq:CLann1pt}) when the boundary field approaches weight zero, $\hP_\beta\to\hP_{(1,1)}$. At $c\leq1$ our bulk--boundary structure constant from Table~\ref{tab:data} tends to a sine in this limit, $\hat R_0(\hP_\alpha,\hP_\beta|\hP_\sigma)\to2\sin(4\pi\hP_\sigma\hP_\alpha)/\hP_\alpha$, and so differs from the disk one-point function $\hat R^{(\text{1-pt})}_0$ which is fixed in our analysis by the consistency of both \eqref{eq:CLann1pt} and \eqref{eq:CLann0pt}. This is expected in timelike Liouville CFT, where the weight-zero limit of a generic vertex operator is not the identity operator, in contrast with the situation at $c\geq25$.
\begin{table*}[t]
\caption{\label{tab:data}Bulk and boundary Liouville data in the bootstrap normalization for $c\ge25$ (middle) and for $c\le1$ (right). Momenta $P,P_i,P_\alpha,P_\beta$ (similarly $\hat{P}_{\cdots}$ for $c\leq1$) label bulk and boundary primaries, $P_{\sigma_i}$ (similarly $\hat{P}_{\sigma_i}$ for $c\leq1$) label boundary conditions. $\rho_0,C_0$ are defined below (\ref{eq:kernels}), the kernels $\F,\Sk$ in (\ref{eq:kernels}), and their Virasoro--Wick rotations $\RF,\RS$ in (\ref{eq:VWR}). Intermediate states are summed with $\int_{i\mathbb R_+}\frac{\mathrm dP}{i}=\int_{i\mathbb R}\frac{\mathrm dP}{2i}$ for $c\ge25$, and with $\int_{i\mathbb R+\epsilon}\frac{\mathrm d\hP}{i}$, $\epsilon>\max|\mathrm{Re}(\hP_{\sigma_a}\pm\hP_{\sigma_b})|$ for $c\le1$; in the latter case every boundary-channel integral (cf.\ (\ref{eq:CLdisk4pt})--(\ref{eq:CLann1pt})) reduces non-trivially to a discrete residue sum over (\ref{eq:Hhatopen}), whose unit multiplicity in the annulus partition function fixes the overall normalization of the right column. The corresponding boundary CFT data therefore are the \emph{residues} of the (appropriate combinations of) meromorphic expressions presented below.}
\begin{ruledtabular}
\begin{tabular}{lll}
quantity & $c\ge25 \ (b\in(0,1])$ & $c\le1\ (\hat{b}\in(0,1])$\\
\colrule
Bulk 2-pt & $\rho^{(b)}_0(P)^{-1}$ & $2\rho^{(\hat{b})}_0(i\hP)/\hP^2$\\
Bulk 3-pt & $C^{(b)}_0(P_1,P_2,P_3)$ & $C^{(\hat{b})}_0(i\hP_1,i\hP_2,i\hP_3)^{-1}$\\
\hline
Bdy 2-pt $S_0(P;P_{\sigma_1},P_{\sigma_2}),\hat{S}_0$ & $(\rho^{(b)}_0(P)C^{(b)}_0(P,P_{\sigma_1},P_{\sigma_2}))^{-1}$ & $\dfrac{\rho^{(\hb)}_0(i\hP_{\sigma_1})\rho^{(\hb)}_0(i\hP_{\sigma_2})\,C^{(\hb)}_0(i\hP,i\hP_{\sigma_1},i\hP_{\sigma_2})}{\sqrt 2\,\hP_{\sigma_1}\hP_{\sigma_2}\hP}$\\[1.2ex]
Bdy 3-pt $(C_0)^{\sigma_3\sigma_2\sigma_1}_{P_3P_2P_1},(\hat{C}_0)$ & $2S_0(P_3;P_{\sigma_1},P_{\sigma_3})\,\F^{(b)}_{P_{\sigma_2},P_3}\sm{P_2}{P_1}{P_{\sigma_3}}{P_{\sigma_1}}$ & $\hat S_0(\hP_3;\hP_{\sigma_1},\hP_{\sigma_3})\,\RF^{(\hb)}_{\hP_{\sigma_2},\hP_3}\sm{\hP_2}{\hP_1}{\hP_{\sigma_3}}{\hP_{\sigma_1}}$\\[0.5ex]
Bulk--bdy 2-pt $R_0(P_\alpha,P_\beta|P_\sigma),\hat{R}_0$ & $2\,\Sk^{(b)}_{P_\sigma,P_\alpha}[P_\beta]/\rho^{(b)}_0(P_\alpha)$ & $\sqrt 2\,\RS^{(\hb)}_{\hP_\sigma,\hP_\alpha}[\hP_\beta]/\hP_\alpha$\\
Disk 1-pt $R^{(\text{1-pt})}_0(P_\alpha|P_\sigma),\hat{R}^{(\text{1-pt})}_0$ & $2\sqrt2\cos(4\pi P_\sigma P_\alpha)/\rho^{(b)}_0(P_\alpha)$ & $4\cos(4\pi\hP_\sigma\hP_\alpha)/\hP_\alpha$\\
\end{tabular}
\end{ruledtabular}
\end{table*}\noindent
\paragraph*{A discrete open spectrum.}
One of the main points of our analysis is that the continuous integration variables associated to the boundary-channel expansions in (\ref{eq:CLdisk4pt})--(\ref{eq:CLann1pt}) do not imply a continuous boundary spectrum. Rather, these integrals provide contour representations of \emph{discrete} boundary-channel expansions, consistent with the annulus
decomposition (\ref{eq:CLann0pt}). To see the mechanism, write a complete boundary-channel integrand as
$f(\hat P)=K(\hat P)\mathcal B_{\hat P}$, where $\mathcal B$ is the full
sphere or torus block (which is an even function of $\hat P$) and $K$ includes the rest of the contributions.
From Table~\ref{tab:data} it is straightforward to see that all such integrands in (\ref{eq:CLdisk4pt})--(\ref{eq:CLann1pt}) satisfy $f(-\hat P)=-f(\hat P)$. Therefore reflection relates the integrals on $\mathrm{Re}\,\hat P=\pm\epsilon$ with opposite signs.
For a regulated meromorphic integrand, closing these two lines gives
\begin{equation}\label{eq:oddstrip}
\begin{aligned}
\int_{i\mathbb R+\epsilon}\frac{d\hat P_{\rm o}}{i}\,f(\hat P_{\rm o})
&=\frac12\lim_{L\to\infty}
  \oint_{\mathcal C_{\epsilon,L}}\frac{d\hat P_{\rm o}}{i}\,f(\hat P_{\rm o})\\
&=\pi\sum_{|\mathrm{Re}\,\hat P_*|<\epsilon}
  \operatorname*{Res}_{\hat P_{\rm o}=\hat P_*}f(\hat P_{\rm o}).
\end{aligned}
\end{equation}
Here
$\mathcal C_{\epsilon,L}$ is the counterclockwise boundary of the rectangle
$|\mathrm{Re}\,\hat P|\leq\epsilon$, $|\mathrm{Im}\,\hat P|\leq L$. The non-trivial point is which residues contribute to the discrete sum. In all sewing constraints (\ref{eq:CLdisk4pt})--(\ref{eq:CLann1pt}), the possible poles of $f(\hat{P})$ fall into three classes: i) poles at
$\hat P=\pm\hat P_{(r,s)}$ (coming from the blocks), ii) poles at $\hat P=\pm\hat P_{(r,-s)}$ (coming from $K$), and iii)
``boundary-dependent'' poles at a special locus
$\hat P=\pm\mathsf P^{e_1e_2}_{(k,l)}$, defined below.
Our main claim is that the first two classes contribute \emph{equal and opposite} residues.
Indeed, denoting the appropriate sphere or torus block residue coefficient by
$\varrho_{r,s}$, we find the common identities
\begin{equation}\label{eq:pairedresidues}
\begin{aligned}
\operatorname*{Res}_{\hat P=\hat P_{(r,s)}}\mathcal B_{\hat P}
&=-\frac{\varrho_{r,s}}{2\hat P_{(r,s)}}\mathcal B_{\hat P_{(r,-s)}},\\
\operatorname*{Res}_{\hat P=\hat P_{(r,-s)}}K(\hat P)
&=\frac{\varrho_{r,s}}{2\hat P_{(r,s)}}K(\hat P_{(r,s)}).
\end{aligned}
\end{equation}
Thus the two residues of $f$ cancel pairwise, leaving a discrete expansion
supported only on the special locus of boundary-dependent momenta
$\hat P=\pm\mathsf P^{e_1e_2}_{(k,l)}$. We note in passing that the closed-channel integrands of (\ref{eq:CLdisk2pt}) and (\ref{eq:CLann1pt}) are odd as well and could likewise be reduced to discrete sums. These, however, do not reflect the closed-string spectrum, which is the continuum of~\cite{Ribault:2015sxa}, already fixed by the bootstrap on the sphere, and exemplified by the closed-channel integrand of (\ref{eq:CLann0pt}) that is even.

This residue structure, together with (\ref{eq:CLann0pt}), leads to the proposed open-string Hilbert space of $c\leq1$ boundary Liouville theory:
\begin{equation}\label{eq:Hhatopen}
\boxed{
\begin{aligned}
&\widehat{\mathcal H}^{\rm open}_{\hat{P}_{\sigma_1},\hat{P}_{\sigma_2}}
=\bigoplus_{e_1,e_2=\pm}\ 
\bigoplus_{k,l=0}^{\infty}\ 
\mathcal M_{\mathsf{P}^{e_1e_2}_{(k,l)}}\,, \\
&\mathsf{P}^{e_1e_2}_{(k,l)}
\equiv e_1\hat{P}_{\sigma_1}
+e_2\hat{P}_{\sigma_2}
+i\Big(\tfrac{\hat{b}+\hat{b}^{-1}}{2}
+k\hat{b}+l\hat{b}^{-1}\Big).
\end{aligned}
}
\end{equation}
Here $\mathcal M_{\mathsf P}$ is the highest-weight representation of the
Virasoro algebra with highest weight $\hat h=\frac{c-1}{24}-\mathsf P^2$, contributing with unit multiplicity. The spectrum remains locally finite
at irrational $\hat b^2$, and its support depends only on the
two boundary conditions flanking the intermediate channel.
The OPE data associated with this discrete spectrum are encoded in the residues of the combinations
of Table~\ref{tab:data} entering the sewing constraints. The full implications of this spectrum will be explored in detail
in the companion paper~\cite{Betzios:2026bcft}.

It is important to mention that, by contrast, the spacelike FZZT open spectrum is a continuum (and analytically continuing the boundary parameters can only add finitely many bound states on top of this continuum~\cite{Teschner:2000md}).

\section{Virasoro--Wick rotation of the Moore--Seiberg relations}
Our numerics verify the sewing constraints (\ref{eq:CLdisk4pt})--(\ref{eq:CLann1pt}) for rational and irrational $\hb^2$ (\SMs). We now explain their analytic origin: a particular continuation of the Moore--Seiberg polynomial relations of $c\geq25$ Liouville theory.

In spacelike Liouville CFT with FZZT boundaries, the $c\geq25$ data of Table~\ref{tab:data} turn the analogues of (\ref{eq:CLdisk4pt})--(\ref{eq:CLann1pt}) into the following relations among the kernels $\F\equiv\F^{(b)}$ and $\Sk\equiv\Sk^{(b)}$:
\par\noindent \emph{Disk four-point crossing} $\Leftrightarrow$ \emph{pentagon equation}:
\begin{equation}\label{eq:pentms}
\begin{aligned}
&\int_{i\mathbb{R}}\!\frac{dP_s}{i}\,\F_{P_o,P_s}\!\sm{P_1}{P_2}{P_u}{P_q}\F_{P_q,P_4}\!\sm{P_s}{P_3}{P_u}{P_5}\F_{P_s,P_t}\!\sm{P_2}{P_3}{P_1}{P_4}\\
&\quad=\F_{P_o,P_4}\!\sm{P_1}{P_t}{P_u}{P_5}\F_{P_q,P_t}\!\sm{P_2}{P_3}{P_o}{P_5}.
\end{aligned}
\end{equation}
\par\noindent \emph{Disk three-point crossing} $\Leftrightarrow$ \emph{torus two-point relation}:
\begin{equation}\label{eq:g1n2ms}
\begin{aligned}
&\Sk_{P',P_2}[P_5]\,\F_{P_1,P_5}\!\sm{P_0'}{P_0}{P'}{P'}=\int_{i\mathbb{R}}\!\frac{dP_3}{i}\frac{dP_4}{i}\,\Sk_{P_1,P_2}[P_3]\\
&\quad\times\F_{P',P_3}\!\sm{P_0'}{P_0}{P_1}{P_1}\F_{P_3,P_4}\!\sm{P_0'}{P_2}{P_0}{P_2}\F_{P_4,P_5}\!\sm{P_0'}{P_0}{P_2}{P_2}\\
&\quad\times e^{i\pi(2h_4-2h_2-h_0-h_0'+\frac{h_3}{2}+\frac{h_5}{2})}.
\end{aligned}
\end{equation}
\par\noindent \emph{Disk two-point crossing} $\Leftrightarrow$ \emph{non-rational Verlinde formula (v1)}:
\begin{equation}\label{eq:VirasoroVerlinde1}
\begin{aligned}
&\int_{i\mathbb{R}}\!\frac{dP_2}{i}\,\rho_0(P_2)C_0(P_2,P_1,P_3)\,\frac{\Sk_{P',P_2}[\mathds{1}]}{\Sk_{\mathds{1},P_2}[\mathds{1}]}\,\F_{P_2,P_0}\!\sm{P_1}{P_1}{P_3}{P_3}\\
&\quad=\rho_0(P_0)C_0(P',P',P_0)\,\frac{\Sk_{P',P_1}[P_0]\,\Sk_{P',P_3}[P_0]}{\Sk_{\mathds{1},P_1}[\mathds{1}]\,\Sk_{\mathds{1},P_3}[\mathds{1}]}.
\end{aligned}
\end{equation}
\par\noindent \emph{Annulus one-point covariance} $\Leftrightarrow$ \emph{non-rational Verlinde formula (v2)}:
\begin{equation}\label{eq:VirasoroVerlinde2}
\int_{i\mathbb{R}}\!\frac{dP_3}{i}\,\F_{P_1,P_3}\!\sm{P_0}{P_3}{P_1}{P_2}\Sk_{P_3,P'}[P_0]=\frac{\Sk_{P_1,P'}[P_0]\,\Sk_{P_2,P'}[\mathds{1}]}{\Sk_{\mathds{1},P'}[\mathds{1}]}.
\end{equation}
Equations (\ref{eq:pentms}), (\ref{eq:g1n2ms}) are standard Moore--Seiberg relations~\cite{Moore:1988qv,Teschner:2003at,Teschner:2008qh,Teschner:2012em,Eberhardt:2023mrq}, and (\ref{eq:VirasoroVerlinde1}), (\ref{eq:VirasoroVerlinde2}) arise from the non-rational Verlinde formula of~\cite{Post:2024itb}.

For $c\leq1$, our main claim is that (\ref{eq:CLdisk4pt})--(\ref{eq:CLann1pt}) are equivalent to ``VWR-ing'' these relations, up to some crucial contour manipulations. We prove this analytically for (\ref{eq:CLdisk4pt}) at rational $\hb^2$; the other cases are discussed in the companion paper~\cite{Betzios:2026bcft}.

Inserting the data of Table~\ref{tab:data}, both sides of (\ref{eq:CLdisk4pt}) become proportional to $\hat S_0(\hP_4;\hP_{\sigma_1},\hP_{\sigma_4})$, and (\ref{eq:CLdisk4pt}) reduces to
\begin{equation}\label{eq:GsGt}
\int_{i\mathbb R+\epsilon}\!\!\tfrac{d\hP}{i}A(\hP)\hat{\mathcal V}_{\eta}(\hP;\hP_{1234})\!=\!\!\int_{i\mathbb R+\epsilon}\!\!\tfrac{d\hP}{i}B(\hP)\hat{\mathcal V}_{1-\eta}(\hP;\hP_{3214}),
\end{equation}
with the odd functions ($\RF\equiv\RF^{(\hb)}$)
\begin{equation}\label{eq:AB}
\begin{aligned}
A(\hP)&\equiv\RF_{\hP_{\sigma_2},\hP}\!\sm{\hP_1}{\hP_2}{\hP_{\sigma_1}}{\hP_{\sigma_3}}\RF_{\hP_{\sigma_3},\hP_4}\!\sm{\hP}{\hP_3}{\hP_{\sigma_1}}{\hP_{\sigma_4}},\\
B(\hP)&\equiv\RF_{\hP_{\sigma_2},\hP_4}\!\sm{\hP_1}{\hP}{\hP_{\sigma_1}}{\hP_{\sigma_4}}\RF_{\hP_{\sigma_3},\hP}\!\sm{\hP_2}{\hP_3}{\hP_{\sigma_2}}{\hP_{\sigma_4}}.
\end{aligned}
\end{equation}
On the other hand, set $b=\hb$ in (\ref{eq:pentms}), rotate all its momenta $P\to i\hP$ and relabel them. By (\ref{eq:VWR}) every kernel becomes a VWR kernel and the prefactors $\hP_t/\hP_s$ cancel, giving (\ref{eq:pentms}) with $\F\to\RF$ and $\int_{i\mathbb R}\frac{dP_s}{i}\to\int_{\mathbb R}d\hP_s$. For $(P_o,P_u,P_q,P_5)=(\hP_{\sigma_2},\hP_{\sigma_1},\hP_{\sigma_3},\hP_{\sigma_4})$, this ``VWR pentagon'' reads
\begin{equation}\label{eq:VWRpent}
\int_{\mathbb R}\!d\hP_s\,A(\hP_s)\,\RF_{\hP_s,\hP_t}\!\sm{\hP_2}{\hP_3}{\hP_1}{\hP_4}=B(\hP_t).
\end{equation}
This equation has the structure of (\ref{eq:GsGt}), but cannot be used directly. First, (\ref{eq:VWRpent}) converges only for $\hP_t$ in a strip $|\mathrm{Im}\,\hP_t|<\xi$ of finite width, whereas (\ref{eq:GsGt}) needs $\hP_t\in i\mathbb R+\epsilon$. Second, $\RF$ does \emph{not} implement crossing: it \emph{annihilates} the $c\leq1$ blocks, $\int_{i\mathbb R+\epsilon}\frac{d\hP_t}{i}\RF_{\hP_s,\hP_t}\hat{\mathcal V}_{1-\eta}(\hP_t;\hP_{3214})=0$.

The way out is to split the meromorphic kernel as
\begin{equation}\label{eq:RFsplit}
\RF=\tfrac{i}{2}\big(\hat\F^{(+)}-\hat\F^{(-)}\big),
\end{equation}
where each $\hat\F^{(\pm)}$ implements crossing of the $c\leq1$ blocks, and the two are exchanged by reflection of either internal momentum, so that $\RF$ is their odd and the physical kernel~\cite{Ribault:2023vqs,Roussillon:2025tmv} $\hat\F=\frac12(\hat\F^{(+)}+\hat\F^{(-)})$ their even combination. This splitting was proven at rational $\hb^2$~\cite{Roussillon:2025tmv}, where $\hat\F^{(\pm)}$ correspond to the two roots of a quadratic polynomial. We expect it to hold for any $\hb^2$, but the irrational case has not been fully established yet (see~\cite{Roussillon:2024wmr}), so our proof builds on the exact rational results. We emphasize that this concerns only the proof: $\RF$ and $\RS$ are the VWR (\ref{eq:VWR}) of the Ponsot--Teschner kernels~\cite{Ponsot:1999uf,Ponsot:2000mt,Teschner:2003at} for any $\hb\in(0,1]$, and hence the data of Table~\ref{tab:data} hold for \emph{any} $\hb^2$ in this range, as our numerics also confirm.

The kernels $\hat\F^{(\pm)}$ have square-root branch points, but these stay in a strip $|\mathrm{Re}\,\hP_s|\leq R_*$, and, multiplied by $A$, $\hat\F^{(+)}$ ($\hat\F^{(-)}$) decays in the lower (upper) right quadrant of the $\hP_s$-plane. Folding the even integrand of (\ref{eq:VWRpent}) onto $[0,\infty)$, splitting $\RF$ only for $\hP_s\geq\delta>R_*$, and rotating the two pieces onto vertical rays (Fig.~\ref{fig:SMcontours}) gives
\begin{equation}\label{eq:Brep}
\begin{aligned}
B(\hP_t)={}&2\int_0^\delta\!d\hP_s\,A(\hP_s)\,\RF_{\hP_s,\hP_t}\\
&+i\!\int_\delta^{\delta-i\infty}\!\!d\hP_s\,A(\hP_s)\,\hat\F^{(+)}_{\hP_s,\hP_t}\\
&-i\!\int_\delta^{\delta+i\infty}\!\!d\hP_s\,A(\hP_s)\,\hat\F^{(-)}_{\hP_s,\hP_t},
\end{aligned}
\end{equation}
which, unlike (\ref{eq:VWRpent}), converges and holds for every $\hP_t\in i\mathbb R+\epsilon$. We multiply (\ref{eq:Brep}) by the $t$-channel block and integrate over $\hP_t$. The first term vanishes by definition of $\RF$, while the two rays turn into $s$-channel blocks and together form the vertical line $\mathrm{Re}\,\hP_s=\delta$: this is the left-hand side of (\ref{eq:GsGt}). The meromorphic kernel $\RF$ thus serves as a structure constant, while the physical kernels $\hat\F^{(\pm)}$ only enter at intermediate steps. The proof assumes two analytic properties of $\hat\F^{(\pm)}$, namely their crossing property on a fixed branch and their growth; both hold exactly at $c=1$, and we have verified them numerically in other cases. Details and numerical checks of every step are given in the \SMs.

\section{Discussion}

We have constructed $c\leq1$ boundary Liouville theory from the analytic bootstrap, with boundary conditions labeled by a continuous Liouville momentum $\hP_\sigma$. We view these as the $c\leq1$ analogues of the usual $c\geq25$ FZZT boundary conditions. Our principal finding is that the open-string spectrum between two such boundaries is discrete, Eq.~(\ref{eq:Hhatopen}). The corresponding OPE data are extracted from appropriate residues of the meromorphic expressions in Table~\ref{tab:data}. We have provided both analytic and numerical evidence for our construction (see \SMs). A more detailed analysis of these data will appear in the companion paper~\cite{Betzios:2026bcft}.

The physical origin of the discrete spectrum remains unclear. A proper analogue of a semiclassical or minisuperspace analysis could test whether this discreteness can be understood from the effective dynamics of the zero mode of the $c\leq1$ Liouville field and the conditions imposed on its wavefunctions, paralleling the analysis of the $c\geq25$ theory. The exact dependence of our expressions on the central charge provides a starting point for investigating such limits. Closely related is the physical interpretation of the contour splittings underlying our analytic arguments (Fig.~\ref{fig:SMcontours}). Beyond establishing convergence and crossing, it would be valuable to determine whether these contours are related to the (upside-down) U-shaped integration contours appearing in recent approaches to $c\leq1$ Liouville theory~\cite{Usciati:2025cdn}.

A natural next step is to identify the analogues of ZZ boundaries within this construction and to clarify their relation to the proposals reviewed in the Introduction. More broadly, our results may shed light on the path-integral formulation of $c\leq1$ Liouville theory on surfaces with and without boundaries, as well as on its applications to 2d cosmological models~\cite{Bautista:2019jau,Anninos:2021ene,Anninos:2024iwf} and to statistical systems such as critical loop models~\cite{Picco:2016ilr,Ikhlef:2015eua,Downing:2025wer} and 2d percolation~\cite{Delfino:2010xm}. Finally, our exact bootstrap data provide concrete benchmarks for probabilistic approaches to $c\leq1$ Liouville theory~\cite{Guillarmou:2023exh,Usciati:2025cdn,Chatterjee:2025yzo,Cercle:2026ihl}, offering an opportunity to connect its algebraic and probabilistic formulations.

\section*{Acknowledgments}
\begin{acknowledgments}
We would like to thank Dio Anninos, Costas Bachas, Scott Collier, Max Downing, Sylvain Ribault, J\"org Teschner, and Yifan Wang for stimulating discussions.\\
P.~B.~acknowledges financial support from the European Research Council (grant BHHQG-101040024), funded by the European Union. This research was supported in part by the Research Foundation -- Flanders (FWO) through a Grant for a Long Stay Abroad (grant number V431626N). I.~T.~is supported by the
European Union (ERC, FUNBOOTS, project number 101043588). Views and opinions expressed are however those of the authors only and do not necessarily reflect those of the European Union or the European Research Council.  Neither the European Union nor the granting authority can be held responsible for them. P.~B.\ and I.~T.\ would also like to thank the Institute for Advanced Study and the organizers of the workshop ``Liouville Theory: Spacelike, Timelike and Beyond''  for hospitality during the course of this work. We acknowledge the use of Claude for proofreading and for assistance with coding and numerics.
\end{acknowledgments}

\bibliography{biblio}

\clearpage
\onecolumngrid
\hypertarget{sup_mat}{}
\section*{Supplemental Material}
\setcounter{equation}{0}\renewcommand{\theequation}{S\arabic{equation}}
\setcounter{table}{0}\renewcommand{\thetable}{S\arabic{table}}
\setcounter{figure}{0}\renewcommand{\thefigure}{S\arabic{figure}}
\subsection*{The DOZZ structure constant and the Virasoro crossing kernels}

Throughout, $b>0$, $Q=b+b^{-1}$, and momenta are as in (\ref{eq:kernels}), with physical values on $i\mathbb R$. The Barnes double gamma function is defined for $\mathrm{Re}\,x>0$ by
\begin{equation}\label{eq:SMGammab}
\log\Gamma_b(x)=\int_0^\infty\frac{dt}{t}\left[\frac{e^{-xt}-e^{-Qt/2}}{(1-e^{-bt})(1-e^{-t/b})}-\frac{(\frac Q2-x)^2}{2}\,e^{-t}-\frac{\frac Q2-x}{t}\right],
\end{equation}
and continued meromorphically by $\Gamma_b(x+b)=\sqrt{2\pi}\,b^{bx-\frac12}\,\Gamma_b(x)/\Gamma(bx)$. It is invariant under $b\to b^{-1}$, has no zeros, and has poles at $x=-mb-nb^{-1}$, $m,n\in\mathbb Z_{\geq0}$. The double sine function is $S_b(x)=\Gamma_b(x)/\Gamma_b(Q-x)$. We write $\Gamma_b(x\pm y)\equiv\Gamma_b(x+y)\,\Gamma_b(x-y)$, and similarly with several signs.

\paragraph*{DOZZ structure constant.} In the normalization of~\cite{Collier:2019weq,Collier:2023cyw},
\begin{equation}\label{eq:SMC0}
C^{(b)}_0(P_1,P_2,P_3)=\frac{\Gamma_b(2Q)}{\sqrt2\,\Gamma_b(Q)^3}\,\frac{\prod_{\pm,\pm,\pm}\Gamma_b\big(\frac Q2\pm P_1\pm P_2\pm P_3\big)}{\prod_{k=1}^3\Gamma_b(Q\pm2P_k)},
\end{equation}
where the numerator contains all eight sign choices. It is symmetric and even in each momentum, and $\F^{(b)}_{\mathds 1,P_1}\sm{P_2}{P_3}{P_2}{P_3}=\frac12\rho^{(b)}_0(P_1)\,C^{(b)}_0(P_1,P_2,P_3)$ with $\rho^{(b)}_0(P)=-4\sqrt2\sin(2\pi bP)\sin(2\pi b^{-1}P)$.

\paragraph*{Modular kernel.} The kernel $\Sk^{(b)}$ of (\ref{eq:kernels}) is Teschner's~\cite{Teschner:2003at}, which in the form of~\cite{Roussillon:2025tmv} reads
\begin{equation}\label{eq:SMS}
\Sk^{(b)}_{P_s,P_t}[P_0]=\frac{\rho^{(b)}_0(P_t)}{2\,S_b(\frac Q2+P_0)}\,\frac{\Gamma_b(Q\pm2P_s)\,\Gamma_b(\frac Q2-P_0\pm2P_t)}{\Gamma_b(Q\pm2P_t)\,\Gamma_b(\frac Q2-P_0\pm2P_s)}\int_{i\mathbb R}\frac{du}{i}\,e^{4\pi iP_su}\prod_{\pm}\frac{S_b\big(u+\frac Q4\pm P_t+\frac{P_0}2\big)}{S_b\big(u+\frac{3Q}4\pm P_t-\frac{P_0}2\big)},
\end{equation}
and converges for $|\mathrm{Re}\,P_s|<\frac12\mathrm{Re}(\frac Q2-P_0)$.

\paragraph*{Fusion kernel.} The kernel $\F^{(b)}$ of (\ref{eq:kernels}) is the Ponsot--Teschner kernel~\cite{Ponsot:1999uf,Ponsot:2000mt}, here in the representation of Teschner and Vartanov~\cite{Teschner:2012em}, written compactly as in~\cite{Roussillon:2025tmv}. Attach the six momenta to the edges $E=\{1,2,3,4,s,t\}$ of a tetrahedron with faces $\mathcal F=\{12s,34s,23t,14t\}$ and vertices $\mathcal V=\{14s,12t,34t,23s\}$, each given by its three edges. For signs $\mu_i=\pm1$, $i\in E$, let $\mu_f=\prod_{i\in f}\mu_i$ for a face or vertex $f$, $\mu_E=\prod_{i\in E}\mu_i$, and $\zeta_f=+1$ if $t\in f$, $\zeta_f=-1$ otherwise. Then
\begin{equation}\label{eq:SMF}
\begin{aligned}
\F^{(b)}_{P_s,P_t}\!\sm{P_2}{P_3}{P_1}{P_4}={}&\frac{\Gamma_b(Q\pm2P_s)}{2\,\Gamma_b(\pm2P_t)}\prod_{f\in\mathcal F}\ \prod_{\mu_f=\zeta_f}\Gamma_b\Big(\frac Q2+\sum_{i\in f}\mu_iP_i\Big)^{\mu_f}\\
&\times\int_{i\mathbb R}\frac{du}{i}\prod_{\substack{\mu\in\{\pm1\}^E\\ \mu_v=1\ \forall v\in\mathcal V}}S_b\Big(u+\frac Q2+\frac Q4\,\mu_E+\frac12\sum_{i\in E}\mu_iP_i\Big)^{-\mu_E}.
\end{aligned}
\end{equation}
The prefactor contains, for each face, the four sign choices on its edges with $\mu_f=\zeta_f$, and the integrand the eight sign choices with $\mu_v=1$ at every vertex. We denote the integral in (\ref{eq:SMF}) by $\mathcal J$.

In (\ref{eq:SMS}) and (\ref{eq:SMF}) the poles of the integrand form increasing sequences $u=u_*+mb+nb^{-1}$ and decreasing sequences $u=u_*-mb-nb^{-1}$, $m,n\in\mathbb Z_{\geq0}$, which for imaginary momenta start at $\mathrm{Re}\,u_*=\frac Q4$ and $-\frac Q4$, respectively. The contour $i\mathbb R$ separates the two kinds; for other momenta it is deformed so that it keeps separating them.

\subsection*{Cyclicity of the boundary three-point structure constants}

In $(C_0)^{\sigma_3\sigma_2\sigma_1}_{P_3P_2P_1}$ at $c\geq25$, the boundary field $\Psi_{P_k}$ sits between the boundary conditions $\sigma_k$ and $\sigma_{k+1}$ ($\sigma_4\equiv\sigma_1$), as in (\ref{eq:CLdisk4pt}). Cyclic symmetry states that the structure constant does not depend on which field is listed first,
\begin{equation}\label{eq:SMcyc}
(C_0)^{\sigma_3\sigma_2\sigma_1}_{P_3P_2P_1}=(C_0)^{\sigma_2\sigma_1\sigma_3}_{P_2P_1P_3},
\end{equation}
and likewise for $(\hat C_0)$ at $c\leq1$. We prove (\ref{eq:SMcyc}) for the data of Table~\ref{tab:data}, at $c\geq25$ and at $c\leq1$, for any $b$ and $\hb$. In this section we write $\sigma_k$ for $P_{\sigma_k}$ and $\hat\sigma_k$ for $\hP_{\sigma_k}$.

\paragraph*{Tetrahedral symmetry.} From $S_b(x+b^{\pm1})=2\sin(\pi b^{\pm1}x)\,S_b(x)$ one finds $\Gamma_b(Q\pm2P)/\Gamma_b(\pm2P)=\rho^{(b)}_0(P)/\sqrt2$. With (\ref{eq:SMC0}), (\ref{eq:SMF}) can then be rewritten as
\begin{equation}\label{eq:SMtetra}
\mathcal T\equiv\frac{\F^{(b)}_{P_s,P_t}\sm{P_2}{P_3}{P_1}{P_4}}{\rho^{(b)}_0(P_t)\,C^{(b)}_0(P_t,P_2,P_3)\,C^{(b)}_0(P_t,P_1,P_4)}=\frac{\Gamma_b(Q)^6}{\sqrt2\,\Gamma_b(2Q)^2}\,\frac{\prod_{e\in E}\Gamma_b(Q\pm2P_e)}{\prod_{f\in\mathcal F}\prod_{\mu_f=-1}\Gamma_b\big(\frac Q2+\sum_{i\in f}\mu_iP_i\big)}\;\mathcal J.
\end{equation}
The right-hand side refers only to the tetrahedron: $\mathcal T$ is invariant under the $24$ permutations of the six momenta that map faces to faces (equivalently, vertices to vertices), and so is the contour prescription of $\mathcal J$.

\paragraph*{The boundary tetrahedron.} The kernel $\F_{\sigma_2,P_3}\sm{P_2}{P_1}{\sigma_3}{\sigma_1}$ of $(C_0)^{\sigma_3\sigma_2\sigma_1}_{P_3P_2P_1}$ has faces $(P_1P_2P_3)$, $(\sigma_1\sigma_2P_1)$, $(\sigma_2\sigma_3P_2)$ and $(\sigma_3\sigma_1P_3)$: the three fields, and each field with its two adjacent boundary conditions. The cyclic relabeling (\ref{eq:SMcyc}) is a rotation of this tetrahedron, and maps the kernel to $\F_{\sigma_1,P_2}\sm{P_1}{P_3}{\sigma_2}{\sigma_3}$. Writing both kernels in the form (\ref{eq:SMtetra}), with the same $\mathcal T$, gives
\begin{equation}\label{eq:SMcycstep}
\frac{\F_{\sigma_2,P_3}\sm{P_2}{P_1}{\sigma_3}{\sigma_1}}{\F_{\sigma_1,P_2}\sm{P_1}{P_3}{\sigma_2}{\sigma_3}}=\frac{\bbPhi^{(b)}_{P_3}[\sigma_1,\sigma_3]}{\bbPhi^{(b)}_{P_2}[\sigma_2,\sigma_3]},\qquad\qquad
\frac{\RF_{\hat\sigma_2,\hP_3}\sm{\hP_2}{\hP_1}{\hat\sigma_3}{\hat\sigma_1}}{\RF_{\hat\sigma_1,\hP_2}\sm{\hP_1}{\hP_3}{\hat\sigma_2}{\hat\sigma_3}}=\frac{\hat\bbPhi^{(\hb)}_{\hP_3}[\hat\sigma_1,\hat\sigma_3]}{\hat\bbPhi^{(\hb)}_{\hP_2}[\hat\sigma_2,\hat\sigma_3]},
\end{equation}
where, at $c\geq25$ and at $c\leq1$ respectively,
\begin{equation}\label{eq:SMPhi}
\bbPhi^{(b)}_{P_3}[P_1,P_2]\equiv\rho^{(b)}_0(P_3)\,C^{(b)}_0(P_1,P_2,P_3),\qquad
\hat\bbPhi^{(\hb)}_{\hP_3}[\hP_1,\hP_2]\equiv4\sqrt2\,\frac{\hat\rho^{(\hb)}_0(\hP_1)\,\hat\rho^{(\hb)}_0(\hP_2)}{\hP_1\hP_2}\,\hP_3\,\hat C^{(\hb)}_0(\hP_1,\hP_2,\hP_3),
\end{equation}
with $\hat\rho^{(\hb)}_0\equiv\hat\rho_0$ and $\hat C^{(\hb)}_0\equiv\hat C_0$ as defined below (\ref{eq:VWR}). Both are symmetric under the exchange of the two momenta in brackets. In the first relation of (\ref{eq:SMcycstep}) the two kernels are $\rho_0(P_3)C_0(P_3,P_2,P_1)C_0(P_3,\sigma_3,\sigma_1)\,\mathcal T$ and $\rho_0(P_2)C_0(P_2,P_1,P_3)C_0(P_2,\sigma_2,\sigma_3)\,\mathcal T$, so $\mathcal T$ and $C_0(P_1,P_2,P_3)$ cancel. For the second relation, (\ref{eq:VWR}) writes $\RF_{\hat\sigma_2,\hP_3}\sm{\hP_2}{\hP_1}{\hat\sigma_3}{\hat\sigma_1}=\frac{\hP_3}{\hat\sigma_2}\F^{(\hb)}_{i\hP_3,i\hat\sigma_2}\sm{i\hP_2}{i\hat\sigma_3}{i\hP_1}{i\hat\sigma_1}$, a kernel on the same tetrahedron with all momenta multiplied by $i$ but with $i\hat\sigma_2$ as its $t$-edge, and similarly $\RF_{\hat\sigma_1,\hP_2}\sm{\hP_1}{\hP_3}{\hat\sigma_2}{\hat\sigma_3}$ with $t$-edge $i\hat\sigma_1$. Applying (\ref{eq:SMtetra}) to both, $\mathcal T$ and $C^{(\hb)}_0(i\hP_1,i\hat\sigma_1,i\hat\sigma_2)$ cancel, and the ratio is
\begin{equation}\label{eq:SMcycstepc1}
\frac{\hP_3\,\hat\sigma_1}{\hP_2\,\hat\sigma_2}\,\frac{\rho^{(\hb)}_0(i\hat\sigma_2)\,C^{(\hb)}_0(i\hP_2,i\hat\sigma_2,i\hat\sigma_3)}{\rho^{(\hb)}_0(i\hat\sigma_1)\,C^{(\hb)}_0(i\hP_3,i\hat\sigma_1,i\hat\sigma_3)}
=\frac{\hat\bbPhi^{(\hb)}_{\hP_3}[\hat\sigma_1,\hat\sigma_3]}{\hat\bbPhi^{(\hb)}_{\hP_2}[\hat\sigma_2,\hat\sigma_3]},
\end{equation}
since $\hat\rho_0(P)=P^2/(2\rho^{(\hb)}_0(iP))$ and $\hat C_0=1/C^{(\hb)}_0(i\,\cdot)$ give
\begin{equation*}
\hat\bbPhi^{(\hb)}_{\hP_3}[\hP_1,\hP_2]=\frac{\sqrt2\,\hP_1\hP_2\hP_3}{\rho^{(\hb)}_0(i\hP_1)\,\rho^{(\hb)}_0(i\hP_2)\,C^{(\hb)}_0(i\hP_1,i\hP_2,i\hP_3)},
\end{equation*}
and the factor $\hat\sigma_3/\rho^{(\hb)}_0(i\hat\sigma_3)$ common to numerator and denominator cancels.

\paragraph*{Cyclicity.} By Table~\ref{tab:data} and the expressions above, the boundary two-point functions are the inverses of (\ref{eq:SMPhi}),
\begin{equation}\label{eq:SMSPhi}
S_0(P;\sigma,\sigma')\,\bbPhi^{(b)}_{P}[\sigma,\sigma']=1,\qquad
\hat S_0(\hP;\hat\sigma,\hat\sigma')\,\hat\bbPhi^{(\hb)}_{\hP}[\hat\sigma,\hat\sigma']=1,
\end{equation}
and their symmetry under the exchange of the two boundary labels is that of the brackets in (\ref{eq:SMPhi}). Hence the right-hand sides of (\ref{eq:SMcycstep}) equal $S_0(P_2;\sigma_2,\sigma_3)/S_0(P_3;\sigma_1,\sigma_3)$ and $\hat S_0(\hP_2;\hat\sigma_2,\hat\sigma_3)/\hat S_0(\hP_3;\hat\sigma_1,\hat\sigma_3)$, and, using this symmetry,
\begin{equation}\label{eq:SMcycproof}
\begin{aligned}
(C_0)^{\sigma_3\sigma_2\sigma_1}_{P_3P_2P_1}&=2S_0(P_3;\sigma_1,\sigma_3)\,\F_{\sigma_2,P_3}\sm{P_2}{P_1}{\sigma_3}{\sigma_1}=2S_0(P_2;\sigma_2,\sigma_3)\,\F_{\sigma_1,P_2}\sm{P_1}{P_3}{\sigma_2}{\sigma_3}=(C_0)^{\sigma_2\sigma_1\sigma_3}_{P_2P_1P_3},\\
(\hat C_0)^{\sigma_3\sigma_2\sigma_1}_{\hP_3\hP_2\hP_1}&=\hat S_0(\hP_3;\hat\sigma_1,\hat\sigma_3)\,\RF_{\hat\sigma_2,\hP_3}\sm{\hP_2}{\hP_1}{\hat\sigma_3}{\hat\sigma_1}=\hat S_0(\hP_2;\hat\sigma_2,\hat\sigma_3)\,\RF_{\hat\sigma_1,\hP_2}\sm{\hP_1}{\hP_3}{\hat\sigma_2}{\hat\sigma_3}=(\hat C_0)^{\sigma_2\sigma_1\sigma_3}_{\hP_2\hP_1\hP_3}.\qquad\blacksquare
\end{aligned}
\end{equation}
Conversely, for a vertex of the form $\hat S_0\,\RF$, (\ref{eq:SMcycstep}) shows that cyclicity holds if and only if $\hat S_0(\hP;\hat\sigma,\hat\sigma')\,\hat\bbPhi^{(\hb)}_{\hP}[\hat\sigma,\hat\sigma']$ depends only on its last label $\hat\sigma'$, the boundary condition $\hat\sigma_3$ common to both sides of (\ref{eq:SMcycstep}); if $\hat S_0$ is symmetric in its labels, this product is then a constant, so cyclicity fixes $\hat S_0\propto1/\hat\bbPhi^{(\hb)}$ up to normalization. We have checked (\ref{eq:SMtetra})--(\ref{eq:SMcycproof}) numerically for generic complex momenta, at $b^2=\frac23$ and $b=0.8$ and at $\hb^2=\frac12,\frac23,\frac{3-\sqrt5}2$, to $10^{-12}$.

\subsection*{Proof of the annulus open-closed duality (\ref{eq:CLann0pt})}

We prove (\ref{eq:CLann0pt}) for any $\hb\in(0,1]$, any $\tau$ in the upper half-plane, any $\epsilon>0$ and arbitrary complex boundary labels, which we write as $\hP_{\sigma_a}=iv_a$, $a=1,2$. We use $q=e^{2\pi i\tau}$, $\hat\chi_P(\tau)=q^{-P^2}/\eta(\tau)$, and
\begin{equation}
p_{r,s}\equiv\tfrac12\big(r\hb+s\hb^{-1}\big),\qquad\text{so that}\qquad \hP_{(r,-s)}=i\,p_{r,s}.
\end{equation}

\paragraph*{Closed-channel density.}
With $\hat\rho_0(P)=P^2/(2\rho^{(\hb)}_0(iP))$, $\rho^{(\hb)}_0(iP)=4\sqrt2\,\sinh(2\pi\hb P)\sinh(2\pi P/\hb)$ and $\hat R_0^{(\text{1-pt})}(P|\hP_\sigma)=4\cos(4\pi\hP_\sigma P)/P$ (Table~\ref{tab:data}), the integrand of the right-hand side of (\ref{eq:CLann0pt}) is
\begin{equation}\label{eq:SMann-density}
\hat\rho_0(P)\,\hat R_0^{(\text{1-pt})}(P|\hP_{\sigma_1})\,\hat R_0^{(\text{1-pt})}(P|\hP_{\sigma_2})
=\frac{\sqrt2\,\cos(4\pi\hP_{\sigma_1}P)\cos(4\pi\hP_{\sigma_2}P)}{\sinh(2\pi\hb P)\,\sinh(2\pi P/\hb)} .
\end{equation}
It is even in $P$, and all of its poles lie on $i\mathbb R$. We set $P=iy$, so that the contour $P\in i\mathbb R+\epsilon$ becomes $y\in\mathbb R-i\epsilon$ and $dP/i=dy$. With $\theta\equiv2\pi\hb y$, $\phi\equiv2\pi y/\hb$, $-P^2=y^2$ and $\eta(-1/\tau)=\sqrt{-i\tau}\,\eta(\tau)$, the right-hand side of (\ref{eq:CLann0pt}) becomes
\begin{equation}\label{eq:SMann-Z}
Z_{\rm closed}=-\frac{\sqrt2}{\sqrt{-i\tau}\,\eta(\tau)}\int_{\mathbb R-i\epsilon}\!dy\;
\frac{\cos(4\pi v_1y)\cos(4\pi v_2y)}{\sin\theta\,\sin\phi}\;e^{-2\pi i y^2/\tau}.
\end{equation}

\paragraph*{Expansion on the shifted contour.}
On the contour $\mathrm{Im}\,y=-\epsilon<0$, so $|e^{-2i\theta}|=e^{-4\pi\hb\epsilon}<1$ and $|e^{-2i\phi}|=e^{-4\pi\epsilon/\hb}<1$. The geometric series $1/\sin\theta=2i\sum_{r\,\text{odd}\geq1}e^{-ir\theta}$ and its analogue for $\phi$ then give
\begin{equation}\label{eq:SMann-expansion}
\frac{1}{\sin\theta\,\sin\phi}=-4\sum_{r,s\ \text{odd}\ \geq1}e^{-4\pi i\,p_{r,s}\,y},
\end{equation}
which converges absolutely and uniformly on the contour, since $|e^{-4\pi ip_{r,s}y}|=e^{-4\pi\epsilon\,p_{r,s}}$. On the real line itself (\ref{eq:SMann-expansion}) would only hold in the Abel-summed sense; the shift by $\epsilon$ is what makes it absolutely convergent. We insert (\ref{eq:SMann-expansion}) and $\cos(4\pi v_1y)\cos(4\pi v_2y)=\frac14\sum_{e_1,e_2=\pm}e^{4\pi i(e_1v_1+e_2v_2)y}$ into (\ref{eq:SMann-Z}). We may exchange sum and integral, because the integrand decays like a Gaussian along the contour ($\mathrm{Re}(i/\tau)=\mathrm{Im}\,\tau/|\tau|^2>0$). This gives
\begin{equation}\label{eq:SMann-sum}
Z_{\rm closed}=\frac{\sqrt2}{\sqrt{-i\tau}\,\eta(\tau)}\sum_{e_1,e_2=\pm}\ \sum_{r,s\ \text{odd}\ \geq1}\ \int_{\mathbb R-i\epsilon}\!dy\;e^{4\pi i\kappa y-2\pi iy^2/\tau},
\qquad \kappa\equiv e_1v_1+e_2v_2-p_{r,s}.
\end{equation}

\paragraph*{Gaussian integrals.}
Each integrand in (\ref{eq:SMann-sum}) is entire in $y$ and decays like a Gaussian throughout the strip, so each contour can be shifted back to $\mathbb R$. Then
\begin{equation}
\int_{\mathbb R}dy\;e^{4\pi i\kappa y-2\pi iy^2/\tau}=\sqrt{-i\tau/2}\;q^{\kappa^2},
\end{equation}
and the constants combine to one, $\sqrt2\,\sqrt{-i\tau/2}/\sqrt{-i\tau}=1$. Relabeling $(e_1,e_2)\to(-e_1,-e_2)$ gives
\begin{equation}\label{eq:SMann-open}
Z_{\rm closed}=\frac{1}{\eta(\tau)}\sum_{e_1,e_2=\pm}\ \sum_{r,s\ \text{odd}\ \geq1}q^{\,(p_{r,s}+e_1v_1+e_2v_2)^2}.
\end{equation}

\paragraph*{Open channel.}
With $(r,s)=(2k+1,2l+1)$, $k,l\geq0$, the momenta (\ref{eq:Hhatopen}) read $\mathsf P^{e_1e_2}_{(k,l)}=i\big(e_1v_1+e_2v_2+p_{2k+1,2l+1}\big)$. Hence $-\big(\mathsf P^{e_1e_2}_{(k,l)}\big)^2=(p_{r,s}+e_1v_1+e_2v_2)^2$, and each term of (\ref{eq:SMann-open}) is exactly $\hat\chi_{\mathsf P^{e_1e_2}_{(k,l)}}(\tau)$, with unit coefficient. This is the left-hand side of (\ref{eq:CLann0pt}). \hfill$\square$

\paragraph*{Remarks.}
(i) The result does not depend on $\epsilon>0$. For $\epsilon<0$ the same steps, with the expansion in $e^{+i\theta}$, $e^{+i\phi}$, give the same answer, since the density (\ref{eq:SMann-density}) is even. The condition $\epsilon>\max|\mathrm{Re}(\hP_{\sigma_a}\pm\hP_{\sigma_b})|$ in Table~\ref{tab:data} is needed only for the open-channel integrals of (\ref{eq:CLdisk4pt})--(\ref{eq:CLann1pt}).
(ii) The unit multiplicity in (\ref{eq:CLann0pt}) fixes the normalization of $\hat R_0^{(\text{1-pt})}$. At rational $\hb^2$, distinct $(k,l)$ can give the same $p_{2k+1,2l+1}$; the corresponding characters then add up to integer multiplicities.
(iii) The discreteness of the open spectrum can be traced to the closed-channel density: on the real $y$ line it is $\propto1/(\sin\theta\sin\phi)$, whose Fourier spectrum is the discrete set $\{p_{r,s}\}$, cf.~(\ref{eq:SMann-expansion}). At $c\geq25$ the same steps lead to $1/\big(\sinh(2\pi by)\sinh(2\pi y/b)\big)$, whose Fourier transform is continuous, and hence to the FZZT open-string continuum~\cite{Teschner:2000md}.
(iv) We have also checked (\ref{eq:CLann0pt}) numerically, with agreement to $10^{-12}$, for several values of $\hb^2$, $\epsilon$ and $\tau$, including complex boundary labels.

\subsection*{Numerical checks of the discrete boundary channels}

\paragraph*{Lattice sums.}
For imaginary boundary labels all poles of the integrands of (\ref{eq:CLdisk4pt})--(\ref{eq:CLann1pt}) lie on $i\mathbb R$. By (\ref{eq:oddstrip}) every boundary-channel integral then equals the discrete sum
\begin{equation}\label{eq:SMlatticesum}
\Sigma=2\pi\sum_{\hP_*\in\Lambda}\ \operatorname*{Res}_{\hP=\hP_*}f(\hP),
\end{equation}
where $f$ is the full integrand (structure constants times conformal block) and $\Lambda$ is the set of momenta $\mathsf P^{e_1e_2}_{(k,l)}$ of (\ref{eq:Hhatopen}) with $\mathrm{Im}\,\hP_*>0$, built from the two labels that flank the channel. The Kac poles are not included, since they cancel in pairs by (\ref{eq:pairedresidues}). At irrational $\hb^2$ the Kac poles $\pm\hP_{(r,s)}$ are dense on $i\mathbb R$; the argument of (\ref{eq:oddstrip})--(\ref{eq:pairedresidues}) is then applied to the blocks truncated at level $N$, which have finitely many such poles ($rs\le N$), the remaining poles of $K$ at $\pm\hP_{(r,-s)}$, $rs>N$, contributing only beyond the truncation order. When the two flanking labels are equal (both channels of (\ref{eq:CLdisk3pt}) and the boundary channel of (\ref{eq:CLdisk2pt})), the points with $e_1=-e_2$ lie at the null-descendant momenta $\hP_{(r,-s)}$. Their residues are Kac terms that drop out together with those of the blocks, and the sum runs over $e_1=e_2$ only; we have checked that these terms alone reproduce the line integrals. Each residue is computed as a trapezoidal sum on a circle of 24 points, with radius at most $0.3$ times the distance to the nearest other lattice or Kac point. At rational $\hb^2$ coincident lattice points give poles of higher order, which this captures automatically. All points with $\mathrm{Im}\,\hP_*\leq6$ are included; the last ones contribute less than $2\cdot10^{-10}$ of the sum, and less than $10^{-20}$ except in (\ref{eq:CLdisk3pt}).

\paragraph*{What is compared.}
Tables~\ref{tab:SM3} and \ref{tab:SM456} below list the relative difference between the two sides of (\ref{eq:CLdisk4pt})--(\ref{eq:CLann1pt}). In (\ref{eq:CLdisk4pt}) and (\ref{eq:CLdisk3pt}) both sides are lattice sums (\ref{eq:SMlatticesum}). In (\ref{eq:CLdisk2pt}) and (\ref{eq:CLann1pt}) the boundary side is the lattice sum and the bulk side, whose spectrum is continuous, is the line integral on $\epsilon+i\mathbb R$ with $\epsilon=0.3$ (Simpson's rule, $|\mathrm{Im}\,\hP|\leq4.5$). The relative differences do not depend on the normalization of the boundary fields, since both sides of every constraint scale in the same way.

\paragraph*{Parameters.}
We consider five rational values of $\hb^2$ and two irrational ones, $\hb^2=1/\sqrt2$ ($c=13-9\sqrt2$) and $\hb^2=\frac{3-\sqrt5}{2}$ ($c=-5$). The external momenta are the same for all central charges. The labels are chosen so that the low-lying lattice points stay away from the Kac points. For rational $\hb^2$ they are at least $0.038$ from every Kac point; for the two irrational values, where the null-descendant momenta $\hP_{(r,-s)}$ are dense, they are at least $0.012$ from the Kac points $\hP_{(r,\pm s)}$ with $rs\leq12$. The labels for the irrational rows are given in brackets.
(\ref{eq:CLdisk4pt}): $\hP_j=iu_j$, $u=(0.13,0.21,0.17,0.29)$, and $\hP_{\sigma_j}=iv_j$, $v=(0.182,0.300,0.063,0.238)$ [$v=(0.119,0.367,0.249,0.353)$].
(\ref{eq:CLdisk3pt}): $\hP_\alpha=0.13i$, $\hP_1=0.21i$, $\hP_2=0.17i$, $(\hP_{\sigma_1},\hP_{\sigma_2})=(0.05i,0.375i)$ [$(0.11i,0.1675i)$].
(\ref{eq:CLdisk2pt}): $\hP_{\alpha_1}=0.13i$, $\hP_{\alpha_2}=0.21i$, $\hP_\sigma=0.05i$ [$0.11i$].
(\ref{eq:CLann1pt}): $\hP_\beta=0.17i$, $(\hP_{\sigma_1},\hP_{\sigma_2})=(0.065i,0.18i)$ [$(0.277i,0.057i)$].
The fusion kernel is evaluated in closed form at rational $\hb^2$~\cite{Roussillon:2025tmv}, and at irrational $\hb^2$ as the integral (\ref{eq:SMF}) (with the residues of the poles that cross the contour when the momenta have real parts), with $\Gamma_b$ from (\ref{eq:SMGammab}); at rational $\hb^2$ the two agree to $10^{-11}$. The modular kernel is obtained from the fusion kernel at Liouville parameter $\sqrt2\,b$~\cite{Hadasz:2009sw,Roussillon:2024wmr},
\begin{equation}\label{eq:SMSfromF}
\Sk^{(b)}_{P_s,P_t}[P_0]=\sqrt2\;256^{\,P_s^2-P_t^2}\;\F^{(\sqrt2b)}_{\sqrt2P_s,\sqrt2P_t}\!\sm{P_0/\sqrt2}{b/(2\sqrt2)}{b/(2\sqrt2)}{b/(2\sqrt2)},
\end{equation}
at $b=\hb$, with the fusion kernel on the right-hand side evaluated as above; (\ref{eq:SMSfromF}) agrees with the integral (\ref{eq:SMS}) to $10^{-11}$. The sphere blocks are computed in the elliptic representation to order $q^{12}$, and the torus blocks to order $q^8$.

\paragraph*{Disk with one bulk and two boundary fields.}
In (\ref{eq:CLdisk3pt}) the bulk field is at $z=e^{i\varphi}$, the boundary fields $\Psi_1$ and $\Psi_2$ are at $0$ and $\infty$, and $\sigma_2$ ($\sigma_1$) is on the positive (negative) real axis.

\emph{Meaning of $\eta$ and $\eta'$.} Both sides of (\ref{eq:CLdisk3pt}) are expansions of one and the same function of $\varphi\in(0,\pi)$, the correlator $\langle\Psi_2(\infty)\,V_{\hP_\alpha}(e^{i\varphi})\,\Psi_1(0)\rangle$: the unprimed side in the channel where the bulk field approaches the boundary $\sigma_2$ ($\varphi\to0$), the primed side in the channel where it approaches $\sigma_1$ ($\varphi\to\pi$). Both are evaluated at the same bulk point, so $\eta$ and $\eta'$ are the same number, $\eta=\eta'=1-e^{-2i\varphi}$, and (\ref{eq:CLdisk3pt}) holds for each $\varphi$. The prime only indicates on which branch the multivalued functions are taken. Each side is defined near its own boundary, where it is fixed by the bulk--boundary expansion (with the principal branch of all powers there), and is continued from there in $\varphi$. As $\varphi$ runs from $0$ to $\pi$, $\eta$ runs once along the circle $|\eta-1|=1$ traced by the doubled bulk points, from $\eta=0$ (with $\eta\simeq2i\varphi$) around $\eta=1$ back to $\eta=0$ (with $\eta\simeq-2i(\pi-\varphi)$). The two sides therefore reach a given $\eta$ from opposite sides of the branch point $\eta=1$ of the blocks. The same applies to the prefactors: $\arg(1-\eta)=-2\varphi$ when continued from $\varphi=0$, and $2\pi-2\varphi$ when continued from $\varphi=\pi$, so that
\begin{equation}\label{eq:SMphases}
(1-\eta)^{\hat p}=e^{-i\varphi A},\qquad(1-\eta')^{\hat p}=e^{i(\pi-\varphi)A},\qquad A\equiv2\hat p=2\hh_\alpha+\hh_1-\hh_2,
\end{equation}
and the two sides carry the relative phase $e^{-i\pi A}$, in addition to the inner phases. By contrast, $\arg\eta=\frac\pi2-\varphi$ stays in $(-\frac\pi2,\frac\pi2)$ for all $\varphi\in(0,\pi)$, whichever end it is continued from. A factor such as $\eta^{\hh_1-\hh_2}$ would therefore take the same value on both sides and drop out of (\ref{eq:CLdisk3pt}), which is why it does not appear in $\phi_\eta$ and $\phi'_{\eta'}$. Finally, the phases $e^{\mp\frac{i\pi}{2}\hh_{\hP}}$ in the integrands, together with $e^{\pm i\pi\hh_\alpha}$ in $\phi_\eta$ and $\phi'_{\eta'}$, make the leading term of each block real and positive near its own boundary, $e^{\mp\frac{i\pi}{2}(\hh_{\hP}-2\hh_\alpha)}\,\eta^{\hh_{\hP}-2\hh_\alpha}\simeq(2\,\mathrm{Im}\,z)^{\hh_{\hP}-2\hh_\alpha}$, as in the bulk--boundary expansion.

\emph{Partial waves.} For the numerics we write the integrands of (\ref{eq:CLdisk3pt}), without the densities, in terms of the modular parameter $\tau$ of $\eta=\lambda(\tau)$:
\begin{equation}\label{eq:SMwaves}
\begin{aligned}
W_{\sigma_2}(\hP)&=e^{-i\varphi A}\,e^{-\frac{i\pi}{2}(\hh_{\hP}-2\hh_\alpha)}\ \hat{\mathcal V}(\hP;\tau_+),&\qquad \tau_+&=\tfrac12+iT(\varphi),\\
W_{\sigma_1}(\hP)&=e^{i(\pi-\varphi) A}\,e^{+\frac{i\pi}{2}(\hh_{\hP}-2\hh_\alpha)}\ \hat{\mathcal V}(\hP;\tau_-),&\qquad \tau_-&=-\tfrac12+iT(\pi-\varphi),
\end{aligned}
\end{equation}
with $A$ as in (\ref{eq:SMphases}), so that $W_{\sigma_2}=\phi_\eta\,e^{-\frac{i\pi}{2}\hh_{\hP}}\hat{\mathcal V}_\eta$ and $W_{\sigma_1}=\phi'_{\eta'}\,e^{\frac{i\pi}{2}\hh_{\hP}}\hat{\mathcal V}_{\eta'}$. Here $\hat{\mathcal V}(\hP;\tau)$ is the sphere block with $\hh_\alpha,\hh_\alpha,\hh_1,\hh_2$ at $0,\lambda(\tau),1,\infty$ ($\lambda=\vartheta_2^4/\vartheta_3^4$), continued in $\tau$ from $i\infty$, and $T(\varphi)>0$ is fixed by $\lambda\big(\tfrac12+iT(\varphi)\big)=1-e^{-2i\varphi}$. Since $\lambda(-\bar\tau)=\overline{\lambda(\tau)}$, also $\lambda(\tau_-)=1-e^{-2i\varphi}$: $\tau_+$ and $\tau_-$ describe the same $\eta$, reached along the two paths. The line $\mathrm{Re}\,\tau=\pm\frac12$ is mapped onto the circle $|\lambda-1|=1$ described by the doubled bulk points, so $T(\tfrac\pi2)=\tfrac12$. The two sides of (\ref{eq:CLdisk3pt}) are $\int\frac{d\hP}{i}$ of the densities shown there times $W_{\sigma_2}$ and $W_{\sigma_1}$, respectively; at $c=-2$ the corresponding line integrals agree to $10^{-9}$ for $\varphi=\frac{5\pi}{12},\frac{7\pi}{12}$. The accuracy of the entries for (\ref{eq:CLdisk3pt}) is limited by the level-12 blocks at the comparatively large nome $|e^{i\pi\tau_\pm}|\approx0.17$--$0.25$ of this configuration.

\begin{table}[!ht]
\caption{\label{tab:SM3}Boundary four-point crossing (\ref{eq:CLdisk4pt}): relative difference $|\Sigma_s-\Sigma_t|/|\Sigma_s|$ between the lattice sums (\ref{eq:SMlatticesum}) of the $s$- and $t$-channels, for several central charges and cross-ratios $\eta$.}
\begin{ruledtabular}
\begin{tabular}{llccccc}
$c$ & $\hb^2$ & $\eta=0.2$ & $\eta=0.35$ & $\eta=0.5$ & $\eta=0.65$ & $\eta=0.8$\\
\colrule
$1$ & $1$ & $3\cdot10^{-9}$ & $1\cdot10^{-11}$ & $1\cdot10^{-12}$ & $2\cdot10^{-12}$ & $3\cdot10^{-12}$\\
$13-9\sqrt2$ & $1/\sqrt2$ & $6\cdot10^{-13}$ & $5\cdot10^{-13}$ & $4\cdot10^{-13}$ & $4\cdot10^{-13}$ & $8\cdot10^{-13}$\\
$0$ & $\tfrac23$ & $2\cdot10^{-12}$ & $2\cdot10^{-12}$ & $1\cdot10^{-12}$ & $2\cdot10^{-12}$ & $2\cdot10^{-12}$\\
$-2$ & $\tfrac12$ & $1\cdot10^{-12}$ & $3\cdot10^{-12}$ & $3\cdot10^{-12}$ & $3\cdot10^{-12}$ & $4\cdot10^{-12}$\\
$-\tfrac{22}{5}$ & $\tfrac25$ & $8\cdot10^{-12}$ & $6\cdot10^{-12}$ & $4\cdot10^{-12}$ & $3\cdot10^{-12}$ & $3\cdot10^{-12}$\\
$-5$ & $\tfrac{3-\sqrt5}{2}$ & $5\cdot10^{-12}$ & $2\cdot10^{-12}$ & $1\cdot10^{-12}$ & $3\cdot10^{-14}$ & $2\cdot10^{-12}$\\
$-7$ & $\tfrac13$ & $4\cdot10^{-11}$ & $2\cdot10^{-11}$ & $2\cdot10^{-11}$ & $1\cdot10^{-11}$ & $8\cdot10^{-12}$\\
\end{tabular}
\end{ruledtabular}
\end{table}

\begin{table}[!ht]
\caption{\label{tab:SM456}Relative difference $|L-R|/|L|$ between the two sides of (\ref{eq:CLdisk3pt}) (both sides lattice sums), and of (\ref{eq:CLdisk2pt}) and (\ref{eq:CLann1pt}) (boundary channel: lattice sum; bulk channel: line integral), as functions of the bulk angle $\varphi$, the cross-ratio $x$ and the modulus $\tau$.}
\begin{ruledtabular}
\begin{tabular}{lccccccccc}
 & \multicolumn{3}{c}{(\ref{eq:CLdisk3pt}) $V\Psi\Psi$, $\varphi=$} & \multicolumn{3}{c}{(\ref{eq:CLdisk2pt}) $VV$, $x=$} & \multicolumn{3}{c}{(\ref{eq:CLann1pt}) annulus, $\tau=$}\\
$c$ & $\tfrac{5\pi}{12}$ & $\tfrac{\pi}{2}$ & $\tfrac{7\pi}{12}$ & $0.3$ & $0.5$ & $0.65$ & $0.8i$ & $i$ & $1.3i$\\
\colrule
$1$ & $2\cdot10^{-7}$ & $1\cdot10^{-8}$ & $6\cdot10^{-8}$ & $3\cdot10^{-13}$ & $3\cdot10^{-13}$ & $3\cdot10^{-13}$ & $3\cdot10^{-13}$ & $3\cdot10^{-13}$ & $5\cdot10^{-13}$\\
$13-9\sqrt2$ & $1\cdot10^{-7}$ & $8\cdot10^{-9}$ & $1\cdot10^{-8}$ & $1\cdot10^{-12}$ & $1\cdot10^{-12}$ & $1\cdot10^{-12}$ & $5\cdot10^{-13}$ & $6\cdot10^{-13}$ & $6\cdot10^{-13}$\\
$0$ & $3\cdot10^{-8}$ & $3\cdot10^{-9}$ & $2\cdot10^{-8}$ & $1\cdot10^{-13}$ & $7\cdot10^{-14}$ & $5\cdot10^{-14}$ & $8\cdot10^{-13}$ & $6\cdot10^{-13}$ & $2\cdot10^{-13}$\\
$-2$ & $1\cdot10^{-7}$ & $3\cdot10^{-9}$ & $1\cdot10^{-7}$ & $1\cdot10^{-13}$ & $2\cdot10^{-13}$ & $3\cdot10^{-13}$ & $5\cdot10^{-14}$ & $1\cdot10^{-13}$ & $2\cdot10^{-13}$\\
$-\tfrac{22}{5}$ & $2\cdot10^{-8}$ & $4\cdot10^{-9}$ & $1\cdot10^{-8}$ & $7\cdot10^{-13}$ & $3\cdot10^{-13}$ & $7\cdot10^{-13}$ & $8\cdot10^{-12}$ & $2\cdot10^{-11}$ & $5\cdot10^{-11}$\\
$-5$ & $1\cdot10^{-7}$ & $1\cdot10^{-8}$ & $1\cdot10^{-8}$ & $4\cdot10^{-13}$ & $6\cdot10^{-13}$ & $6\cdot10^{-13}$ & $6\cdot10^{-13}$ & $7\cdot10^{-13}$ & $1\cdot10^{-12}$\\
$-7$ & $2\cdot10^{-8}$ & $6\cdot10^{-10}$ & $6\cdot10^{-9}$ & $9\cdot10^{-13}$ & $6\cdot10^{-13}$ & $6\cdot10^{-13}$ & $2\cdot10^{-13}$ & $4\cdot10^{-13}$ & $2\cdot10^{-12}$\\
\end{tabular}
\end{ruledtabular}
\end{table}

\paragraph*{Results.}
In all cases the two sides agree to the accuracy of the numerics (Tables~\ref{tab:SM3} and \ref{tab:SM456}), at rational and irrational $\hb^2$ alike. The boundary channels of (\ref{eq:CLdisk4pt}), (\ref{eq:CLdisk2pt}) and (\ref{eq:CLann1pt}) reproduce the other side to $10^{-11}$--$10^{-14}$. The few larger entries, and all entries for (\ref{eq:CLdisk3pt}), are limited by the precision of the level-12 blocks at larger nome near Kac points, not by the lattice sums. At irrational $\hb^2$ the lattice points are all distinct and no Kac point sits at $\hP=0$, so these rows test the residue mechanism (\ref{eq:oddstrip})--(\ref{eq:pairedresidues}) without the coincidences of rational central charge.

\subsection*{Boundary four-point crossing from the VWR pentagon}

We prove (\ref{eq:CLdisk4pt}) in the form (\ref{eq:GsGt}) at rational $\hb^2=m/n$, starting from the VWR pentagon (\ref{eq:VWRpent}). We drop hats on momenta, $P_j\equiv\hP_j$, $\sigma_j\equiv\hP_{\sigma_j}$, and write
\begin{equation*}
\mathcal V^s_P\equiv\hat{\mathcal V}_\eta(P;P_{1234}),\quad \mathcal V^t_P\equiv\hat{\mathcal V}_{1-\eta}(P;P_{3214}),\quad \RF_{P_s,P_t}\equiv\RF_{P_s,P_t}\!\sm{P_2}{P_3}{P_1}{P_4},
\end{equation*}
and similarly for $\hat\F^{(\pm)}$. A kernel $\mathbb K$ \emph{fuses} if $\mathcal V^s_{P_s}=\int_{i\mathbb R+\epsilon}\frac{dP_t}{i}\,\mathbb K_{P_s,P_t}\mathcal V^t_{P_t}$. We must show $G_s=G_t$, where $G_s\equiv\int_{i\mathbb R+\epsilon_s}\frac{dP}{i}A\,\mathcal V^s$ and $G_t\equiv\int_{i\mathbb R+\epsilon_t}\frac{dP}{i}B\,\mathcal V^t$.

\paragraph*{Properties.} We take $P_j=iu_j$, $\sigma_j=iv_j$ with $u_j,v_j$ real and generic. (P1) $\RF$ is odd in its internal and even in its external momenta, so $A$ and $B$ are odd while the blocks are even. (P2) All poles lie on $i\mathbb R$: those of the blocks at $P_{(r,s)}$; those of $A$ at the Kac points $\pm P_{(r,-s)}$ and on the lattice $\pm\sigma_1\pm\sigma_3\pm\frac i2(r\hb+s\hb^{-1})$, $r,s$ odd (and similarly for $B$ with $(\sigma_2,\sigma_4)$). In the VWR pentagon (\ref{eq:VWRpent}), the poles of its kernel $\RF_{P_s,P_t}$ as a function of $P_s$ lie, for any $P_t$, at $\pm P_1\pm P_2\pm\frac i2(r\hb+s\hb^{-1})$ and $\pm P_3\pm P_4\pm\frac i2(r\hb+s\hb^{-1})$, $r,s$ odd. (P3) $A\,\mathcal V^s$ and $B\,\mathcal V^t$ decay like Gaussians along vertical lines, so by (P2) $G_s$ and $G_t$ do not depend on $\epsilon_s,\epsilon_t>0$. The VWR pentagon (\ref{eq:VWRpent}) is first derived for real momenta. It continues to imaginary labels and momenta as long as no pole of its integrand crosses $\mathbb R$, i.e.\ for $|u_1|+|u_2|$, $|u_3|+|u_4|$, $|v_1|+|v_3|<\frac12(\hb+\hb^{-1})$, which we assume until the end. It then holds for real $P_t$ near $\epsilon_t$.

\paragraph*{Ingredients.} (I1) \emph{Split.} The closed form of the $c\geq25$ kernel at $b^2=m/n$~\cite{Roussillon:2025tmv} is a difference $\F=T_1-T_2$ of two terms. They are attached to the roots $z_{1,2}=(-\beta\pm\sqrt\Delta)/2\alpha$ of a quadratic equation (the ``quantum fusion polynomial'') $\alpha z^2+\beta z+\gamma=0$ whose coefficients are trigonometric polynomials in the momenta, and $\sqrt\Delta\to-\sqrt\Delta$ maps $(T_1,T_2)\to(-T_2,-T_1)$. Defining $\hat\F^{(+)}\equiv-2i\frac{P_t}{P_s}T_1$ and $\hat\F^{(-)}\equiv-2i\frac{P_t}{P_s}T_2$ at the arguments of (\ref{eq:VWR}) gives (\ref{eq:RFsplit}). Besides the poles (P2), $\hat\F^{(\pm)}$ have square-root branch points at the zeros of $\Delta$. Under a reflection $P_s\to-P_s$ or $P_t\to-P_t$, with the branch of $\sqrt\Delta$ chosen so that $\sqrt\Delta\to-\sqrt\Delta$ (as happens automatically at $c=1$, where $\sqrt\Delta$ is entire and odd), the two kernels are exchanged, $\hat\F^{(\pm)}\to\hat\F^{(\mp)}$; at fixed $\sqrt\Delta$ each would instead change sign, since $T_{1,2}$ are then even in all momenta.

\emph{Lemma 1.} Let $\epsilon_t>0$ and $0<\nu<\epsilon_t$. There is $R<\infty$ such that, for $|\mathrm{Re}\,P_t-\epsilon_t|\leq\nu$, every zero of $\Delta$ in $P_s$ has $|\mathrm{Re}\,P_s|\leq R$.
\emph{Proof.} Let $\mathsf s=\sqrt{mn}$ and $\omega=e^{-2\pi\mathsf sP_s}$. The coefficients $\alpha,\gamma$ of the quantum fusion polynomial are linear in $\omega^{\pm1/2}$, and $\beta$ depends on $P_s$ only through a term $\propto\sinh(2\pi\mathsf sP_t)(\omega^{-1}-\omega)$. Hence $\omega^2\Delta=4\sinh^2(2\pi\mathsf sP_t)(\omega^4+1)+c_2\omega^3+c_0\omega^2+c_{-2}\omega$, with $c_k$ bounded for bounded $\mathrm{Re}\,P_t$, while $|\sinh(2\pi\mathsf sP_t)|\geq\sinh(2\pi\mathsf s(\epsilon_t-\nu))>0$. With $M$ the supremum of $|c_k|/|4\sinh^2(2\pi\mathsf sP_t)|$, a zero with $|\omega|=r>1$ obeys $r^4-1\leq M(r^3+r^2+r)<M\frac{r^4-1}{r-1}$, so $r<1+M$. The same holds for $\omega^{-1}$, and $|\omega|=e^{-2\pi\mathsf s\,\mathrm{Re}\,P_s}$. $\square$

Let $R_*$ be the largest $|\mathrm{Re}\,P_s|$ of these zeros and fix $\delta>R_*$. On $\mathcal D=\{\mathrm{Re}\,P_s\geq\delta\}\times\{|\mathrm{Re}\,P_t-\epsilon_t|<\nu\}$ the pair $(\hat\F^{(+)},\hat\F^{(-)})$ then has two continuous determinations, exchanged by $(\hat\F^{(+)},\hat\F^{(-)})\to(-\hat\F^{(-)},-\hat\F^{(+)})$.

(I2) \emph{Fusion.} On one of them, the physical determination, both pieces fuse, $\int_{i\mathbb R+\epsilon_t}\frac{dP_t}{i}\hat\F^{(\pm)}_{P_s,P_t}\mathcal V^t_{P_t}=\mathcal V^s_{P_s}$ for $\mathrm{Re}\,P_s\geq\delta$; on the other one the right-hand side is $-\mathcal V^s_{P_s}$.
(G) \emph{Growth.} On the same determination, $A\,\hat\F^{(+)}$ decays faster than $|P_s|^{-1}$ in the quadrant $\{\mathrm{Re}\,P_s\geq\delta,\,\mathrm{Im}\,P_s\leq0\}$ and $A\,\hat\F^{(-)}$ in $\{\mathrm{Re}\,P_s\geq\delta,\,\mathrm{Im}\,P_s\geq0\}$; on the rays $\delta\mp iY$ they are bounded by $C e^{-cY}$, $c>0$, locally uniformly in $P_t$; and the double integrals of Step 5 below converge absolutely.

\noindent (I2) and (G) are the two inputs that we do not prove in general: Ref.~\cite{Roussillon:2025tmv} shows that $\hat\F^{(\pm)}$ are crossing kernels but fixes no branch. Both hold exactly at $c=1$ (see below), and we have checked them numerically at $c=-2$ and $0$.

\emph{Lemma 2 (annihilation).} $\int_{i\mathbb R+\epsilon_t}\frac{dP_t}{i}\RF_{P_s,P_t}\mathcal V^t_{P_t}=0$ for all $P_s$ in a connected domain $\mathcal U\supset(0,\delta]\cup\{\mathrm{Re}\,P_s\geq\delta\}$ that avoids the poles of $\RF$.
\emph{Proof.} For $\mathrm{Re}\,P_s\geq\delta$ it is the difference of the two equations in (I2), by (\ref{eq:RFsplit}). The left-hand side is meromorphic on $\mathcal U$, so it vanishes identically. $\square$

\begin{figure}[t]
\centering
\begin{tikzpicture}[scale=1.01]
\fill[gray!25] (-0.13,-2.7) rectangle (0.13,2.7);
\draw[->] (-1.6,0) -- (4.3,0) node[right] {$\mathrm{Re}\,P_s$};
\draw[->] (0,-2.8) -- (0,2.9) node[above] {$\mathrm{Im}\,P_s$};
\foreach \y in {0.75,1.3,1.85,2.4} { \node at (0,\y) {\small$\times$}; \node at (0,-\y) {\small$\times$}; }
\draw[very thick,blue] (0,0) -- (1.3,0);
\node[blue,above] at (0.62,0.02) {\small $2\,\RF$};
\draw[very thick,red,->] (1.3,0) -- (1.3,-1.4);
\draw[very thick,red] (1.3,-1.3) -- (1.3,-2.65);
\node[red,right] at (1.35,-1.6) {\small $+\,i\,\hat{\F}^{(+)}$};
\draw[very thick,green!50!black,->] (1.3,0) -- (1.3,1.4);
\draw[very thick,green!50!black] (1.3,1.3) -- (1.3,2.65);
\node[green!50!black,right] at (1.35,1.6) {\small $-\,i\,\hat{\F}^{(-)}$};
\draw[thick,dashed] (1.3,0) -- (4.1,0);
\draw[->,dashed,red] (3.7,-0.08) arc (0:-70:2.4);
\draw[->,dashed,green!50!black] (3.7,0.08) arc (0:70:2.4);
\node[below right] at (1.32,-0.02) {\small $\delta$};
\node[below left] at (0,0) {\small $0$};
\node[gray!70!black,left] at (-0.15,-2.3) {\small $|\mathrm{Re}\,P_s|\le R_*$};
\node at (3.4,2.4) {$P_s$-plane};
\end{tikzpicture}
\caption{\label{fig:SMcontours}\footnotesize{Contours of the representation (\ref{eq:Brep}) of $B(P_t)$. Using the fact that the integrand is even, the original VWR pentagon integrates $A\,\RF$ over $[0,\infty)$. On $[0,\delta]$ (blue) the meromorphic $\RF$ is kept. On $[\delta,\infty)$ (dashed) $\RF=\frac i2(\hat\F^{(+)}-\hat\F^{(-)})$, and the $\hat\F^{(+)}$ piece is rotated down (red) and the $\hat\F^{(-)}$ piece up (green). The grey strip contains all branch points of $\hat\F^{(\pm)}$ for $P_t$ on the physical line (Lemma~1); crosses are poles on $i\mathbb R$.}}
\end{figure}

\paragraph*{Proof.} \emph{Step 1 (fold).} By (P1), $A(P_s)\RF_{P_s,P_t}$ is even in $P_s$, and it is regular at $P_s=0$. Hence (\ref{eq:VWRpent}) gives $B(P_t)=2\int_0^\delta A\,\RF+2\int_\delta^\infty A\,\RF$ for real $P_t$ near $\epsilon_t$.
\emph{Step 2 (split).} On $[\delta,\infty)\subset\mathcal D$ we insert (\ref{eq:RFsplit}) on the physical determination; by (G) the two pieces converge separately. On $[0,\delta]$, where branch points may sit (even on the real axis), we keep $\RF$.
\emph{Step 3 (rotate).} $A\,\hat\F^{(+)}$ is analytic in $\{\mathrm{Re}\,P_s>\delta,\,\mathrm{Im}\,P_s<0\}$ by (P2) and Lemma~1, and its integral over large arcs vanishes by (G). So its half line can be rotated onto the downward ray $\delta-i[0,\infty)$, and likewise the $\hat\F^{(-)}$ piece onto $\delta+i[0,\infty)$ (Fig.~\ref{fig:SMcontours}). This gives (\ref{eq:Brep}) for real $P_t$ near $\epsilon_t$.
\emph{Step 4 (continue).} Both sides of (\ref{eq:Brep}) are analytic in $P_t$ on the strip $\{|\mathrm{Re}\,P_t-\epsilon_t|<\nu\}$. The left-hand side has its poles on $i\mathbb R$. On the right, the segment term has a compact range and the poles of $\RF$ in $P_t$ lie on $i\mathbb R$; the ray terms meet no branch points (Lemma~1) and converge locally uniformly by (G). The two sides agree on a real interval, hence on the whole strip, and in particular for all $P_t\in i\mathbb R+\epsilon_t$, far outside the strip $|\mathrm{Im}\,P_t|<\xi$ in which (\ref{eq:VWRpent}) converges.
\emph{Step 5 (fuse).} We multiply (\ref{eq:Brep}) by $\mathcal V^t_{P_t}$ and integrate over $i\mathbb R+\epsilon_t$, exchanging the integrations by (G). The segment term vanishes by Lemma~2, and (I2) turns the ray terms into $s$-channel blocks:
\begin{equation}
G_t=i\!\int_\delta^{\delta-i\infty}\!\!dP_s\,A\,\mathcal V^s_{P_s}-i\!\int_\delta^{\delta+i\infty}\!\!dP_s\,A\,\mathcal V^s_{P_s}=\int_{i\mathbb R+\delta}\frac{dP_s}{i}\,A\,\mathcal V^s_{P_s}=G_s,
\end{equation}
where the last step moves the line to $\mathrm{Re}\,P_s=\epsilon_s$ using (P2) and (P3). $\blacksquare$

The final contour never passes through $P_s=0$, so no principal value is needed, even though at rational $\hb^2$ the blocks have a pole there.

\paragraph*{Extensions.} (a) The identity extends by analytic continuation to all imaginary $P_j$ and $\sigma_j$, away from the codimension-one loci where $A$ or $B$ have poles as functions of the momenta and labels. (b) For labels with real parts it holds as long as the contours pass to the right of the open-string poles, $\epsilon_s>|\mathrm{Re}\,\sigma_1|+|\mathrm{Re}\,\sigma_3|$ and $\epsilon_t>|\mathrm{Re}\,\sigma_2|+|\mathrm{Re}\,\sigma_4|$; this is the prescription of Table~\ref{tab:data}. (c) For irrational $\hb^2$ it should follow by continuity in $\hb$; we have not written the uniformity argument, but the irrational rows of Table~\ref{tab:SM3} confirm (\ref{eq:CLdisk4pt}) there directly.

\paragraph*{The $c=1$ example.} At $\hb=1$ with all eight momenta equal to $\frac i4$, every step is elementary. Here $\hat\F^{(\pm)}_{P_s,P_t}=16^{P_t^2-P_s^2}e^{\mp2\pi iP_sP_t}$ are entire, and $A(P)=B(P)=-i\,2^{1/4}16^{P^2}f(P)$ with $f(P)=\sinh^2(\frac{\pi P}{2})/\sinh(2\pi P)$. (In $A$ and $B$ one kernel has $P$ in an external slot; it equals $-\sinh(\frac{\pi P}2)/\sinh(2\pi P)$, not the value $\RF_{P_s,P_t}=16^{P_t^2-P_s^2}\sin(2\pi P_sP_t)$ that holds when all external momenta are $\frac i4$.) The VWR pentagon becomes the self-duality $\int_{\mathbb R}dx\,f(x)\sin(2\pi xp)=f(p)$, which converges only for $|\mathrm{Im}\,p|<\frac12$. Fusion and annihilation reduce to Gaussian integrals, (I2) and (G) hold exactly, and Steps 1--5 prove $\mathcal I(T)=T^{-1/2}\mathcal I(1/T)$ for $\mathcal I(T)=\int_{i\mathbb R+\epsilon}\frac{dP}{i}f(P)e^{\pi TP^2}$, which is (\ref{eq:GsGt}) at $\eta=\lambda(iT)$.

\paragraph*{Numerical checks.} We have checked every step at $c=-2$ and $c=0$ ($\hb^2=\frac12,\frac23$), with $u=(0.13,0.21,0.17,0.29)$, $v=(0.19,0.21,0.24,0.28)$, $\epsilon_t=0.5$, $\delta=0.3$, $\eta=\frac12$, closed-form kernels and level-12 blocks. The physical determination is the one continuous from the principal branch of $\sqrt\Delta$ at $(P_s,P_t)=(\delta,\epsilon_t)$.
\begin{itemize}
\item The VWR pentagon (\ref{eq:VWRpent}) holds to $10^{-10}$ at real $P_t$.
\item The branch points are confined to $|\mathrm{Re}\,P_s|\leq R_*=0.0195$ ($c=-2$) and $0.0035$ ($c=0$).
\item The rotation of Step 3 holds to $10^{-10}$. At the end of the rays ($Y=12$), $|A\,\hat\F^{(\pm)}|\approx2\cdot10^{-18}$ on the ray where it is used, while the other piece has grown to $|A\,\hat\F^{(\mp)}|\approx2\cdot10^{15}$.
\item The representation (\ref{eq:Brep}) holds along $P_t=0.5+iy$ up to $y=2.6$, far beyond $\xi\approx0.6$, to $10^{-9}$ of the $L^1$ norm of its integrands.
\item Annihilation (Lemma~2) holds to $10^{-13}$ of the $L^1$ norm of the integrand, and fusion (I2) holds with sign $+1$ to $3\cdot10^{-8}$ or better.
\item At $c=-2$ the two sides of (\ref{eq:GsGt}) agree to $10^{-12}$ and reproduce the line-integral value of (\ref{eq:CLdisk4pt}).
\end{itemize}

\end{document}